# On the Origin of Chirality in Plasmonic Meta-molecules

Kevin Martens[1], Timon Funck[1], Eva Y. Santiago[2], Alexander O. Govorov[2], Sven Burger[3, 4] and Tim Liedl[1*]

[1]Faculty of Physics, Ludwig-Maximilians-University, Geschwister-Scholl-Platz 1, 80539 Munich, Germany

[2]Department of Physics and Astronomy, Nanoscale and Quantum Phenomena Institute, Ohio University, Athens, Ohio 45701, United States

[3]Zuse Institute Berlin, Takustraße 7, D-14195 Berlin, Germany

[4]JCMwave GmbH, Bolivarallee 22, 14050 Berlin, Germany

* tim.liedl@lmu.de

**Chirality is a fundamental feature in all domains of nature, ranging from particle physics over electromagnetism to chemistry and biology.[1-4] Chiral objects lack a mirror plane and inversion symmetry and therefore cannot be spatially aligned with their mirrored counterpart, their enantiomer. Both natural molecules and artificial chiral nanostructures can be characterized by their light-matter interaction, which is reflected in circular dichroism (CD).[5, 6] Using DNA origami,[7, 8] we assemble model meta-molecules from multiple plasmonic nanoparticles, representing meta-atoms accurately positioned in space.[9-11] This allows us to reconstruct piece by piece the impact of varying macromolecular geometries on their surrounding optical near fields. Next to the emergence of CD signatures in the instance that we architect a third dimension, we design and implement sign flipping signals through addition or removal of single particles in the artificial molecules. Our data and theoretical modelling reveal the hitherto unrecognized phenomenon of chiral plasmonic-dielectric coupling, explaining the intricate electromagnetic interactions within hybrid DNA-based plasmonic nanostructures.**

Most biomolecules are chiral and this property plays a crucial role in molecular recognition and functionality, as these processes often depend on enantiomer-selective activity.[12] CD spectroscopy thus has become a wide-spread method to analyse molecular compounds and proteins in science and industry. But despite much effort, however, we are not yet in the position to reliably predict the CD spectra of a macromolecule nor, the other way around, derive the molecular structure from a recorded CD spectrum.[13] The resonances of the atomic bonds within a molecule determine its absorption spectrum and it is the arrangement of the bonds relative to each other, which defines chirality and thus the CD signal. Synthetic and biological molecules usually exhibit their optical responses in the UV range. Metallic nanostructures or architectures of metallic nanoparticles, in contrast, can be designed to feature signals in the visible spectral range.[14-18] In such structures, the plasmons – collective oscillations of electrons in metals – of the metallic surfaces couple to each other and create resonances that interact with the optical near field. Just as in their molecular counterparts, these resonances within the metallic nanostructure, or meta-molecule, lead to characteristic absorption and CD spectra. By tailoring the electromagnetic fields we can thus learn to understand the optical responses of nanostructures and molecules alike. Along this path, meta-materials with novel optical properties,[19, 20] sensing devices,[21-24] and enantioselective catalysis applications[25, 26] have been studied.

Chiral nanostructures can be built from silica,[27] quantum dots,[28-30] or metallic nanoparticles.[14-18] Among those, plasmonic nanoparticles have proven to be well-suited candidates for artificial chiral nanostructures. While chirally shaped metal nanoparticles can already express strong CD responses,[31, 32] the assembly of achiral nanoparticles into chiral architectures offers significantly more freedom in design. Here, the plasmons of the individual particles couple to each other leading to interactions with the surrounding optical fields that have been studied by many groups.[33] For example, ligand-protected gold and silver particle clusters have shown chiral responses[34, 35] and chiral assemblies have been achieved through several materials, including peptides,[36] polyfluorenes,[37] silica films,[38] oxalamide fibers,[39] proteins[40] or micelles.[41] Chiral arrangements of individual entities have been achieved through electron-beam lithography,[42, 43] however, DNA nanotechnology[44] opened up routes to design nanostructures[45, 46] with tailored optical responses on even smaller lengthscales.[14, 15, 46-50] In particular the DNA origami technique[7, 8] proved to be powerful to position nanoparticles with sub-nanometer accuracy.[11, 51] In DNA origami, a single-stranded scaffold is folded into any desired three-dimensional (3D) shape by a multitude of designed short DNA staple strands.[52] The resulting objects are fully addressable using DNA handle strands that protrude from the structure and can capture any object that is functionalized with the complementary DNA sequence. DNA origami therefore provides a freely customizable molecular framework.

Helical shapes do not only play a fundamental role in biology – most notably the right-handed form of double-stranded DNA and the alpha-helical protein units – helical shapes also offer a step-by-step path to understand space and the exact mechanisms behind chiral optical activity. Theoretical models have shown that geometrical features of pitch-length, number of particles or particles per turn can have significant influences on the CD response.[53-55, 33, 15, 16, 18] In this study we present two types of model meta molecules for simulating and predicting the CD responses of gold nanoparticle helices that are assembled particle by particle on a DNA origami scaffold. Our findings contribute to a detailed understanding about the origins of molecular and plasmonic CD and we elucidate in this context the role of dielectric materials in the vicinity of plasmonic arrangements.

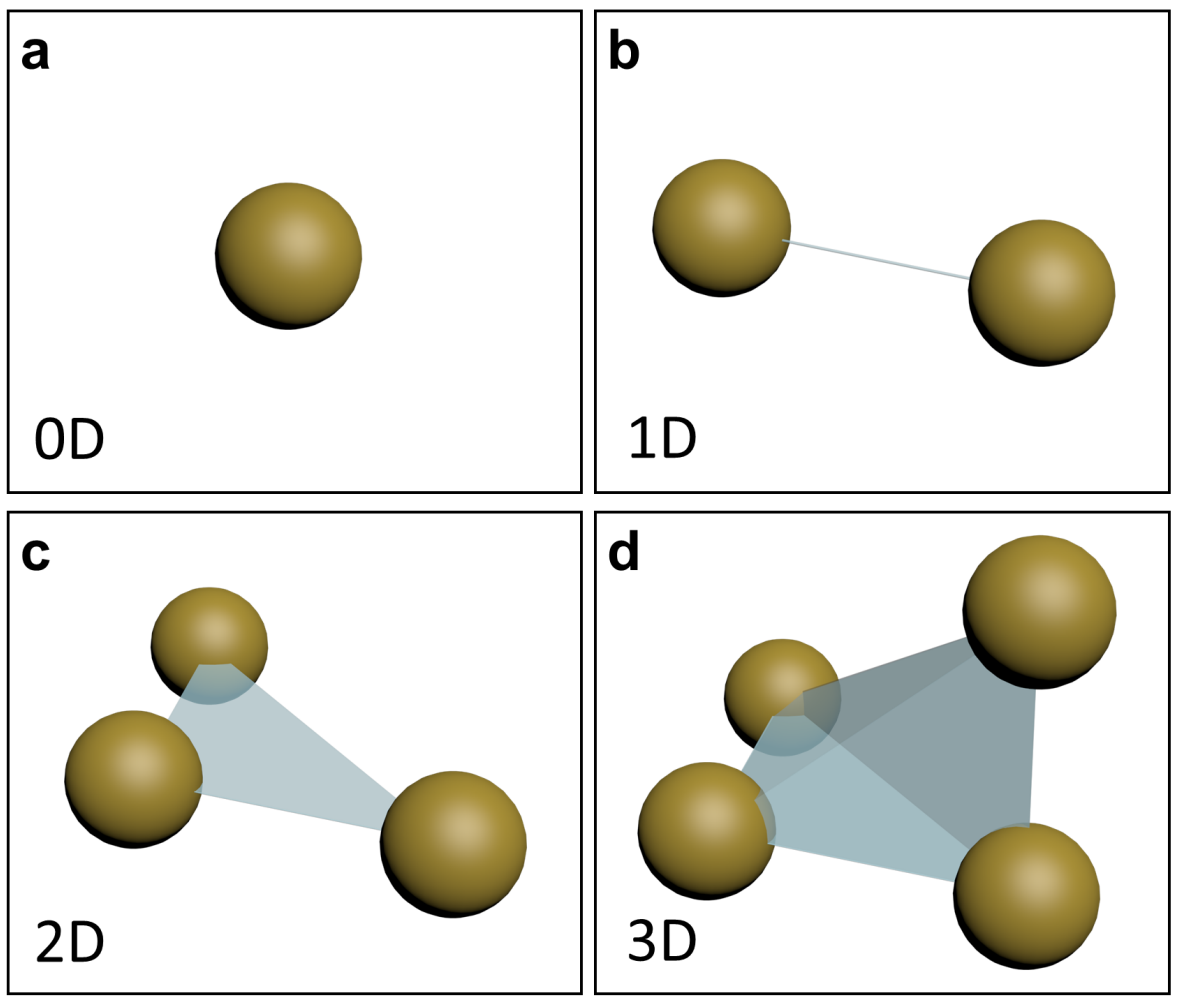

**Figure 1 | Considerations on spheres in space. a**, One sphere in a 0-dimensinal geometry in space. **b**, Two spheres in a 1-dimensional geometry defining a line in space. **c**, Three spheres in a 2-dimensional geometry defining a plane in space. **d**, Four spheres in a 3-dimesional geometry in space, enabling chirality.

Approaching the concept of chirality through an arrangement of spherical particles in space, one may consider how the given number of particles determines the highest possible dimension of any attainable geometrical configuration. While any single sphere ($S_1$) is coercively a 0-dimensional assembly (Fig. 1a), two spheres, $S_1$ and $S_2$, always define a line L connecting them. Such a 1-dimensional assembly has an infinite number of mirror planes through L as well as one mirror plane perpendicular to L at half way between the particles (Fig. 1b). When adding a third sphere $S_3$, two possibilities arise. Either $S_3$ will be located on L, leaving the geometry 1-dimensional or it will be located at any other arbitrary position in space, expanding the structure to the second dimension. In the latter case, $S_1$, $S_2$, and $S_3$ will form a plane P, which will automatically function as the structure's imperative mirror plane (Fig. 1c). An additional mirror plane that is perpendicular to P may arise if the three particles form an isosceles triangle. Finally, when the fourth particle $S_4$ is added (and not placed on P), the structure becomes three-dimensional (Fig. 1d). If this particle is positioned such that no mirror plane exists anymore the structure is chiral, as is our example in Figure 1d. Note that by adding further particles, a previously chiral structure can become achiral or, vice versa, an achiral structure can become chiral.

Here we explore the onset of chirality by assembling unit-by-unit two types of nanoparticle helices around a DNA origami trunk. One helix type contains up to seven 40 nm gold particles with an offset of 29 nm and a rotation around the helical axis of 90° per particle (design details are given in Supplementary Information note 1). We term this helix Large Pitch Helix, LPH, as the pitch will be crucial to understand the various CD responses of helices assembled from 4, 5 and 6 particles. The second helix is termed Small Pitch Helix, SPH, and can host up to six 30 nm spherical gold nanoparticles with an offset of 11 nm. This geometry leads to the particular feature that the fifth particle is located right above the first one at a distance very similar to the distances between the first and the second particle, or second and third etc.. All gold particles are functionalized with thiol-modified DNA sequences, the 'anchors', which are complementary to DNA sequences, the 'handles', protruding from the surface of a multi-helix DNA origami bundle, a 24-helix bundle in case of the LPH and a 48-helix bundle for the SPH. We prepared samples containing varying numbers of particles, always starting with a single particle at one end of the trunk (assembly and purification is described in Supplementary Information note 3). Figure 2 displays schematic drawings and TEM images of all samples.

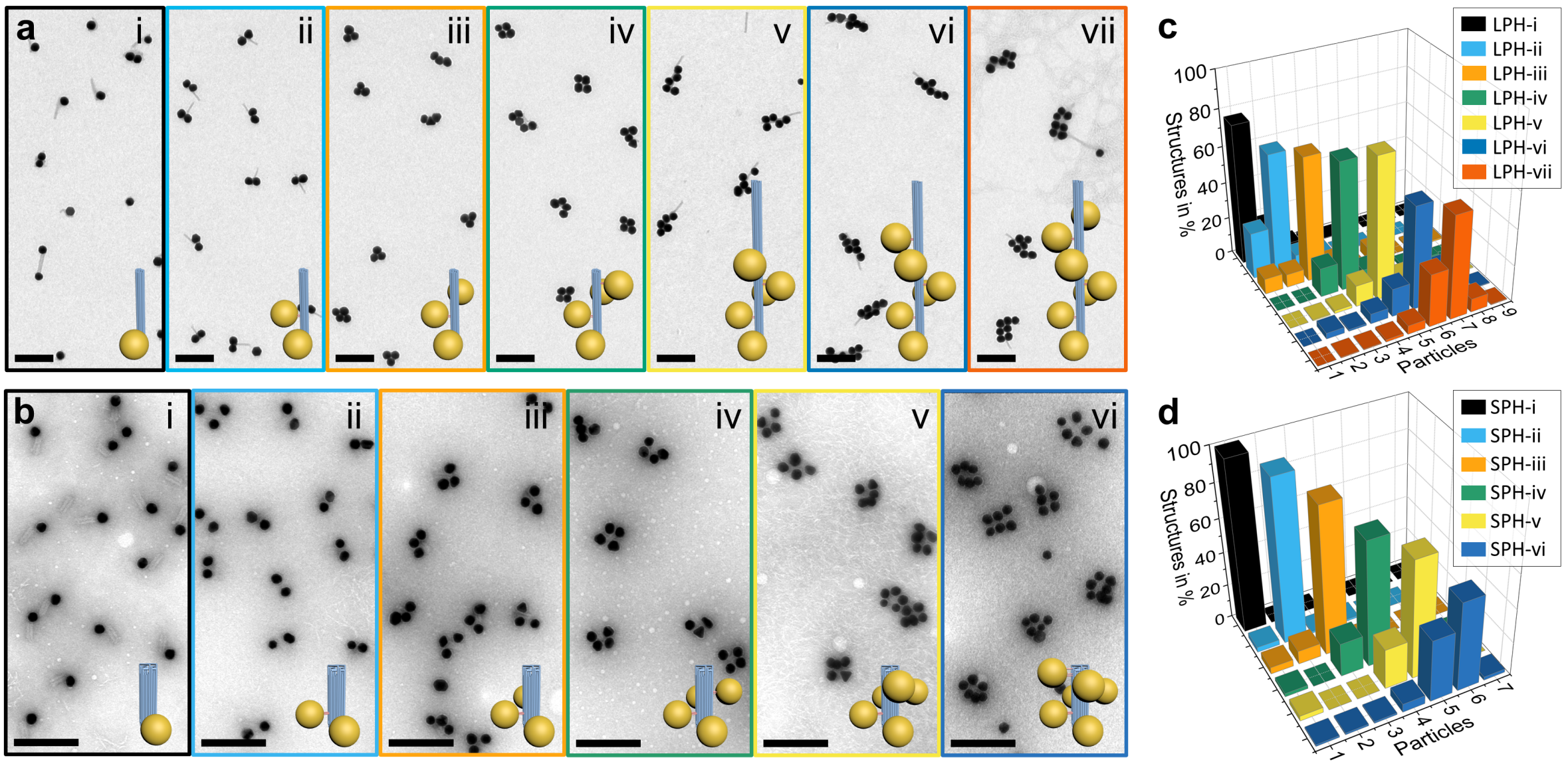

**Figure 2 | Gold Nanohelix Synthesis. a**, Large Pitch Helices (LPH) with different numbers of 40 nm particles attached (LPH-i to LPH-vii) and **b**, Small Pitch Helices (SPH) with different numbers of 30 nm particles attached (SPH-i to SPH-vi), displayed in electron micrographs and 3D models. Scale bar: 200 nm. **c**, Distribution of assemblies in the samples LPH-i to LPH-vii and **d**, distribution of assemblies in the samples SPH-i to SPH-vi. Structures assembled as designed pose the majority for all samples. Structures with one particle less attached are increasingly common in samples with higher particle numbers, whereas structures with more particles then intended are less common for LPH and even negligible for SPH. Approximately 250 individual assemblies per sample were studied.

We achieved satisfactory yields for correctly assembled LPH-i to LPH-vii architectures of 76%, 65%, 68%, 70%, 77%, 56% and 56%, respectively (Fig. 2 a, b and Supplementary Information note 3). Most misassemblies are the result of nanoparticles binding non-specifically to the ends of the multi-helix DNA bundles, leading to fractions of 7%-18% of the structures carrying one ore more extra particles. We therefore added a design feature to the 48-helix bundle of the SPH, where the otherwise accessible ends of the DNA helices are protected by short DNA duplexes crossing the top and bottom of the origami trunk (details on design and synthesis are given in Supplementary Information note 1-3). With this modification we eliminated the unintended binding events resulting in lower fractions of structures carrying more particles then intended (< 3%) for the SPH. This way we achieved yields for assembled SPH-i to SPH-vi architectures of 97%, 96%, 87%, 74%, 70%, and 53%, respectively (Fig. 2 c, d).

Metallic nanoparticle-based meta-molecules are optically most active near their plasmonic resonance frequency, resulting in the highest absorbance as well as CD responses around this frequency. For left-handed objects, as used in this study, typically bisignate peak-dip signals are observed and we obtain exactly such signals with peaks around 524 nm and dips around 558 nm for both our minimal chiral meta-molecules, i.e. the helices containing exactly four particles (Fig. 3 a, b and Supplementary Information note 4). Also, just as expected, we observe no or only negligible CD responses for samples carrying one or two particles, i.e. for SPH-i, SPH-ii, LPH-i and LPH-ii. Surprisingly, the three particle meta-molecules show a CD response around 535 nm and the signal of the large pitch sample (LPH-iii) even shows a dip-peak-shaped spectrum. Numerical simulations for gold "helices" with only 3 particles – obviously 3 particles only form a plane and not a helix – do not show any CD signal (Supplementary Information note 5). If, however, the simulations account for the DNA origami, which here is approximated as a dielectric cylinder penetrating the plane formed by three spheres in an oblique angle, a CD response emerges (Fig. 3 c, d). These augmented simulations yield CD signals not only for the LPH-iii and SPH-iii samples but also noticeable resonances for the LPH-ii and SPH-ii assemblies, where the symmetry is equally broken. Remarkably, the five-particle helices show distinct behavior depending on their pitch. In the case of the LPH, every added particle augments the chiral shape of the meta-molecule, leading to a noticeable increase of the signal (Fig. 3a). Our simulations corroborate this observation (Fig. 3c). For the short-pitch helix the situation is different. Here the fifth particle is very close to the first particle. So instead of increasing the CD signal, the plasmon-plasmon interactions between particle $S_1$ and $S_5$, which are located right on top of each other, induce a right-handed geometry within the structure, which effectively leads to annihilation of the original left-handed signal. This effect leads to SPH-v showing close to no CD response (Figure 3b), which is accurately reflected in the simulated spectra (Figure 3d). The addition of a sixth particle rescues the signal, both in experiment and theory. Figure 3e displays the peak CD intensities for all experimental samples and the corresponding simulation results showing excellent agreement. Also the absorbance peak intensities of the experiments match those of our simulations (Fig. 3f). Typical for assemblies that exhibit plasmonic coupling, the spectra shift into the red with increasing numbers of particles.

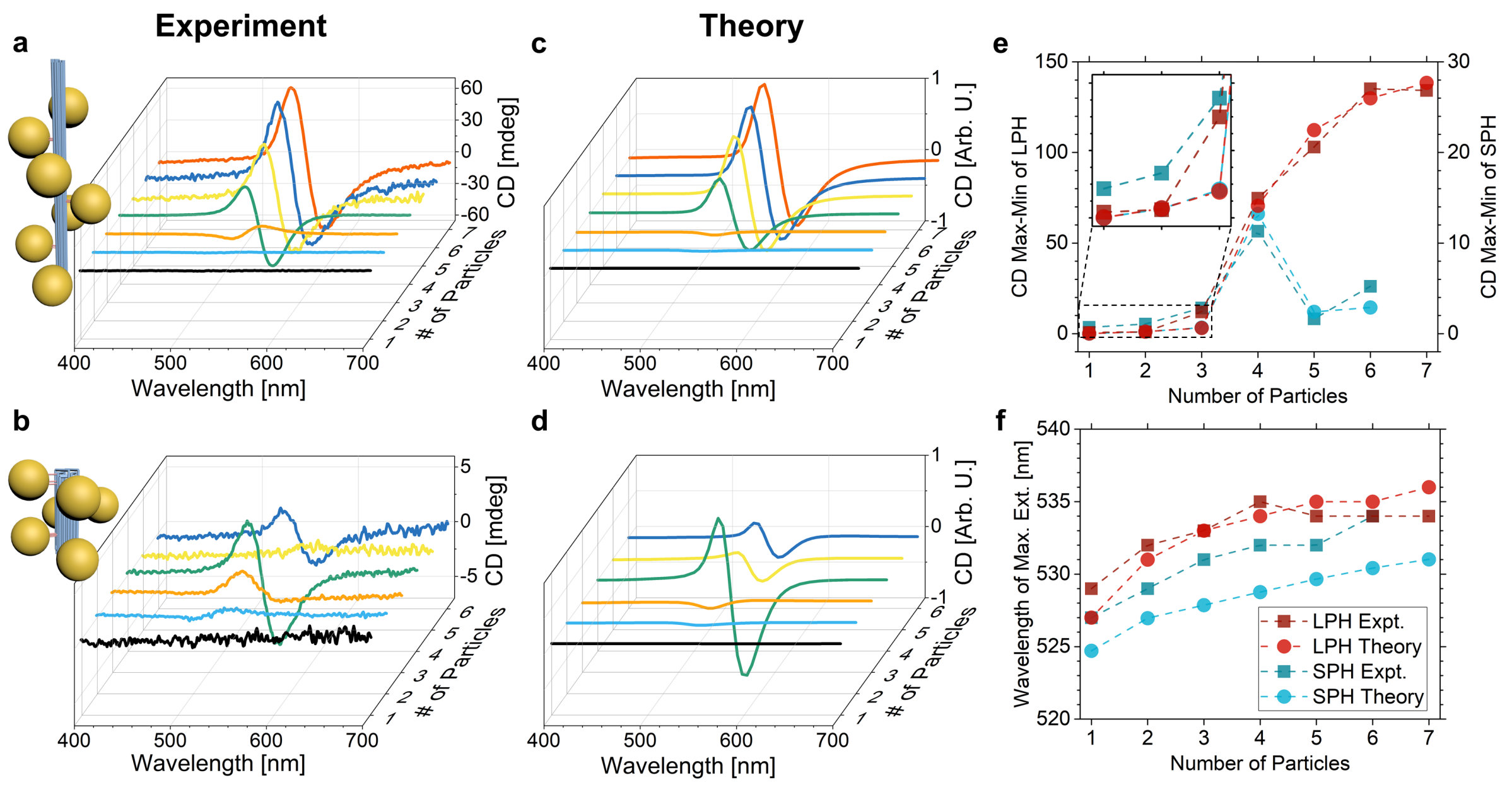

**Figure 3 | Emergence of CD in experiment and theory. a**, CD spectra of samples LPH-i through LPH-vii and **b**, samples SPH-i through SPH-vi, all normalized by the maximum extinction value for each sample. **c**, Simulated CD signal of structures LPH-i through LPH-vii and **d**, structures SPH-i through SPH-vi. The scale is matched to the strength of the experimental CD. **e**, Differences between the maximum and minimum CD response and **f**, wavelength of the peak extinction for LPH-i through LPH-vii (red) as well as SPH-i through SPH-vi (blue) in experiment (squares) and theory (circles).

Figure 4 shows the energy densities of the electric field *u* color-coded on the surfaces of the metallic nanoparticles and the dielectric DNA cylinder. Neighboring particles exhibit plasmonic coupling, which manifests itself as increased energy densities, or hot spots, wherever their surfaces come close to each other. Notably, where the nanoparticles are close to the dielectric surface, we also observe the formation of weak hot spots. Hence, a discernible CD signal emerges already for the cases of three nanoparticles (LPH: Fig. 4a, SPH: Fig. 4b) owing to the dielectric cylinder protruding through the nanoparticle plane in an oblique angle, leading to an overall chiral assembly. The shape of this CD signal differs from the bisignate spectra that are usually observed for helical assemblies but describes a single dip that is rather typical for right-handed assemblies. Four particles define a full helical turn and thus clear, bisgnate CD signals of left-handed helices appear both for the 4-particle LPH and the 4-particle SPH. In the 5-particle case no coupling between the 1$^{st}$ and 5$^{th}$ particle arises in the long pitch helix (Fig 4a), thus the structure gains chiral features and the signal increases further for every added particle. Strikingly, in the case of the short pitch helix, a new hot spot between the 1$^{st}$ and the 5$^{th}$ particle becomes apparent (Fig. 3b), which drastically changes the near field of the incoming light and results in an almost complete breakdown of the CD signal.

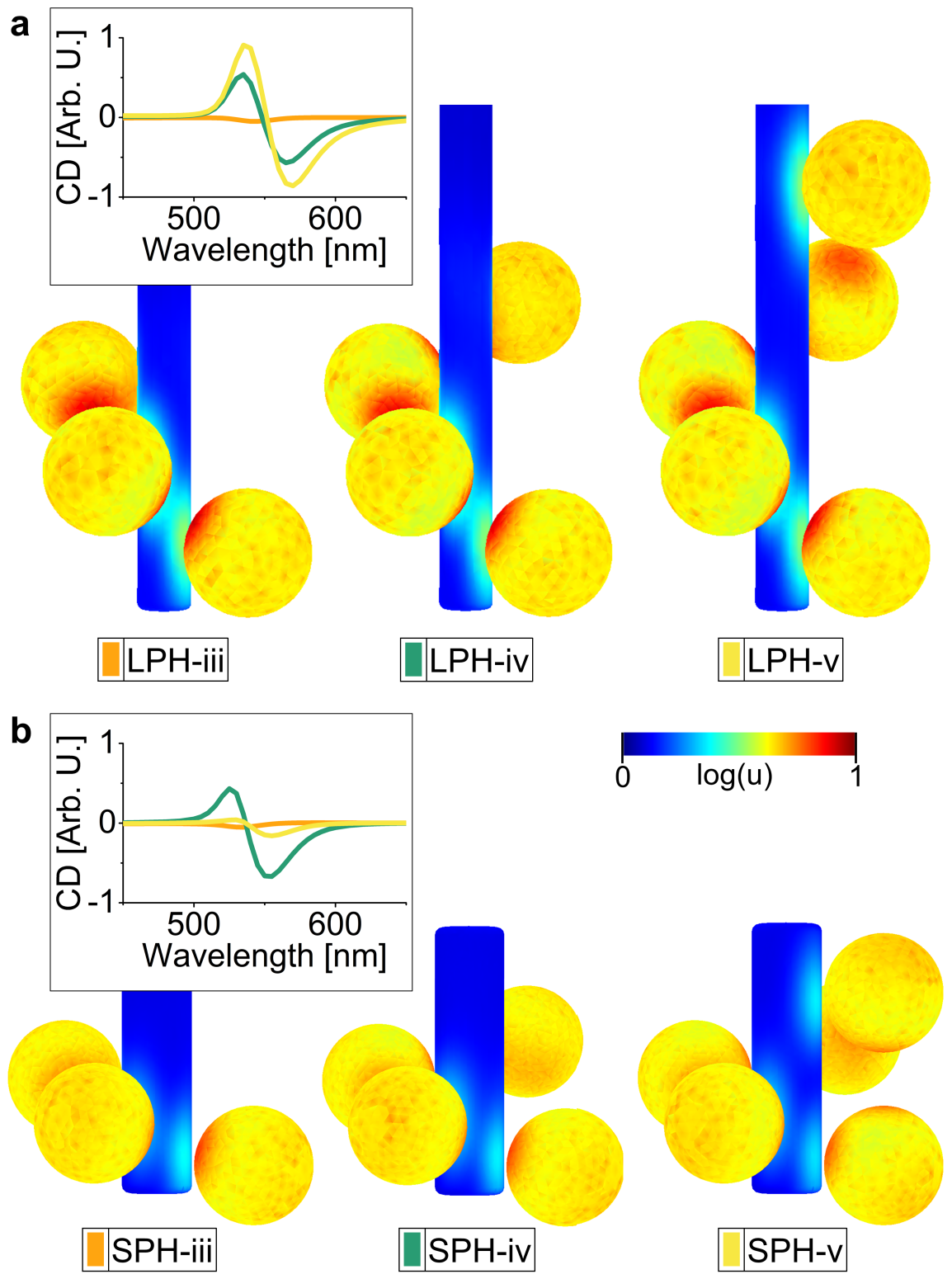

**Figure 4 | Electric field energy density and formation of hot spots.** The simulated optical field energy density *u* visualized in pseudocolors [a.u.] on the surfaces of the gold nanoparticles and the dielectric DNA rods for three critical cases each for the long (**a**) and short (**b**) pitch helices. Insets show simulated CD spectra corresponding to each case.

Using DNA origami and plasmonic particles we are able to design plasmonic meta-atoms and sculpt light matter interactions through carefully considering the geometrical features of two bio-assembled model systems with nanometer precision. Our simulated CD responses in multi-angle illumination settings, reflecting the situation of meta-molecules tumbling in solution, explain the experimentally observed spectral details, in accordance with the computed near field distributions revealing resonant hot spots not only between the metallic particles but also between particles and the DNA. This model, incorporating the dielectric response of the DNA scaffold, adds an additional layer of understanding to the complex analysis of CD spectra of both artificial chiral assemblies and natural molecules.

**Acknowledgements:** Kevin Martens, Timon Funck and Tim Liedl are grateful for financial support through the ERC Consolidator Grant 818635 DNA Funs and the Deutsche Forschungsgemeinschaft (DFG, German Research Foundation) through the SFB1032 (Project A6) and the Cluster of Excellence e-conversion. Sven Burger acknowledges funding by the Deutsche Forschungsgemeinschaft under Germany´s Excellence Strategy – The Berlin Mathematics Research Center MATH+ (EXC-2046/1, project ID: 390685689) and by the German Federal Ministry of Education and Research (BMBF Forschungscampus MODAL, project number 05M20ZBM). Alexander O. Govorov thanks the Berlin Mathematics Research Center MATH+ and Zuse-Institut Berlin for their generous support.

Author Contributions: KM, TF AOG, SB and TL designed the research, KM, TF and TL designed the nanostructures. KM produced the structures and performed the experiments. EYS, AOG and SB performed theoretical calculations. All authors wrote the manuscript.

Author Information: Reprints and permissions information is available. The authors declare no competing financial or non-financial interests. Correspondence and requests for materials should be addressed to TL (tim.liedl@lmu.de).

**Supplementary Information**

# On the Origin of Chirality in Plasmonic Meta-molecules

*Kevin Martens*[1]*, Timon Funck*[1]*,* Eva Y. Santiago[2]*, Alexander O. Govorov*[2]*, Sven Burger*[3 4,] *and Tim Liedl*[1*]

[1]*Faculty of Physics, Ludwig-Maximilians-University, Geschwister-Scholl-Platz 1, D-80539 Munich, Germany*

[2]*Department of Physics and Astronomy, Nanoscale and Quantum Phenomena Institute, Ohio University, Athens, Ohio 45701, United States*

[3]*Zuse Institute Berlin, Takustraße 7, D-14195 Berlin, Germany*

[4]*Department of Physics and Astronomy, Nanoscale and Quantum Phenomena Institute, Ohio University, Athens, Ohio 45701, United States*

[4]*JCMwave GmbH, Bolivarallee 22, D-14050 Berlin, Germany*

* *tim.liedl@lmu.de*

## Supplementary Methods:

DNA scaffold strands (p8064) were prepared following previously described procedures.[1, 2] Unmodified staple strands (purification: desalting) were purchased from Eurofins MWG. Thiol-modified strands (purification: HPLC) were purchased from Biomers. Uranyl formate for negative TEM staining was purchased from Polysciences, Inc.. Spherical gold nanoparticles were purchased from BBI Solutions. Other chemicals were purchased from CarlRoth and Sigma-Aldrich.

# Supplementary Note 1: DNA Origami and Nanohelix Design

The Large Pitch Helix (LPH) was formed around a 24 helix bundle (24 HB) DNA origami trunk. This origami was synthesized with an 8064 nt scaffold strand and 156 core staple strands. For a monomer 24 HB, used for LPH-i through LPH-iv, $C_4$ endcap strand extensions were added to each side (45 in total), which prevent any further attachment by leaving 4 cytosine bases protruding from the origami structure. For the larger gold helices (LPH-v through LPH-vii) a dimeric 24 HB was synthesized. For this, only one side is folded with $C_4$ endcap strands while the other side uses dimerization staples, which leave gaps and protruding strands that can attach to the complementary dimerization staples of second 24 HB origami (See Supplementary Figure 3). Gold nanoparticles can be attached to protruding handle strands depicted in red.

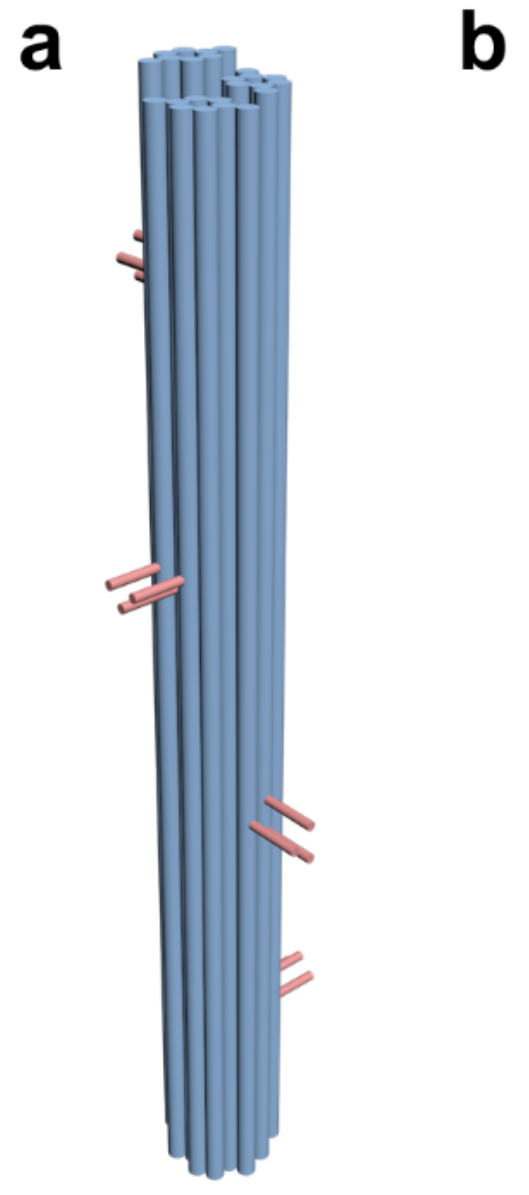

**Supplementary Figure 1: 3D Model of the DNA 24 HB origami structure**

**a**, Monomer structure and **b**, dimer structure with individual halves depicted in blue and green (cylinders represent DNA helices) and handles for nanoparticle attachment depicted in red.

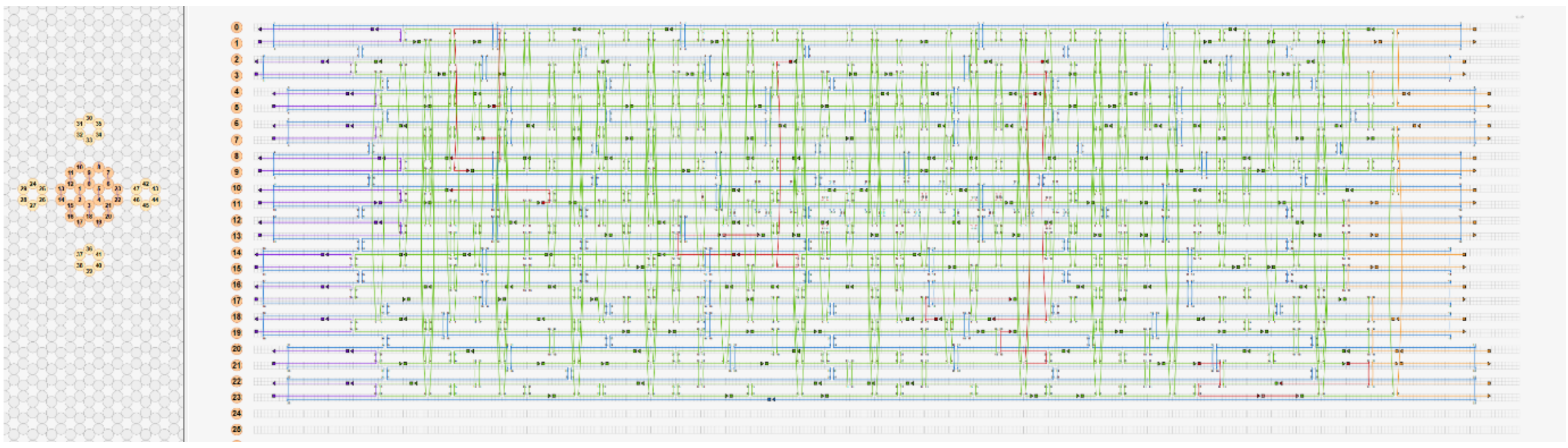

**Supplementary Figure 2: Cadnano design of the 24 HB monomer**

The scaffold strand is depicted in blue, C4 endcaps are depicted in purple (left) and orange (right). Core staples are depicted in green as well as the handle strands for the NPs in red.

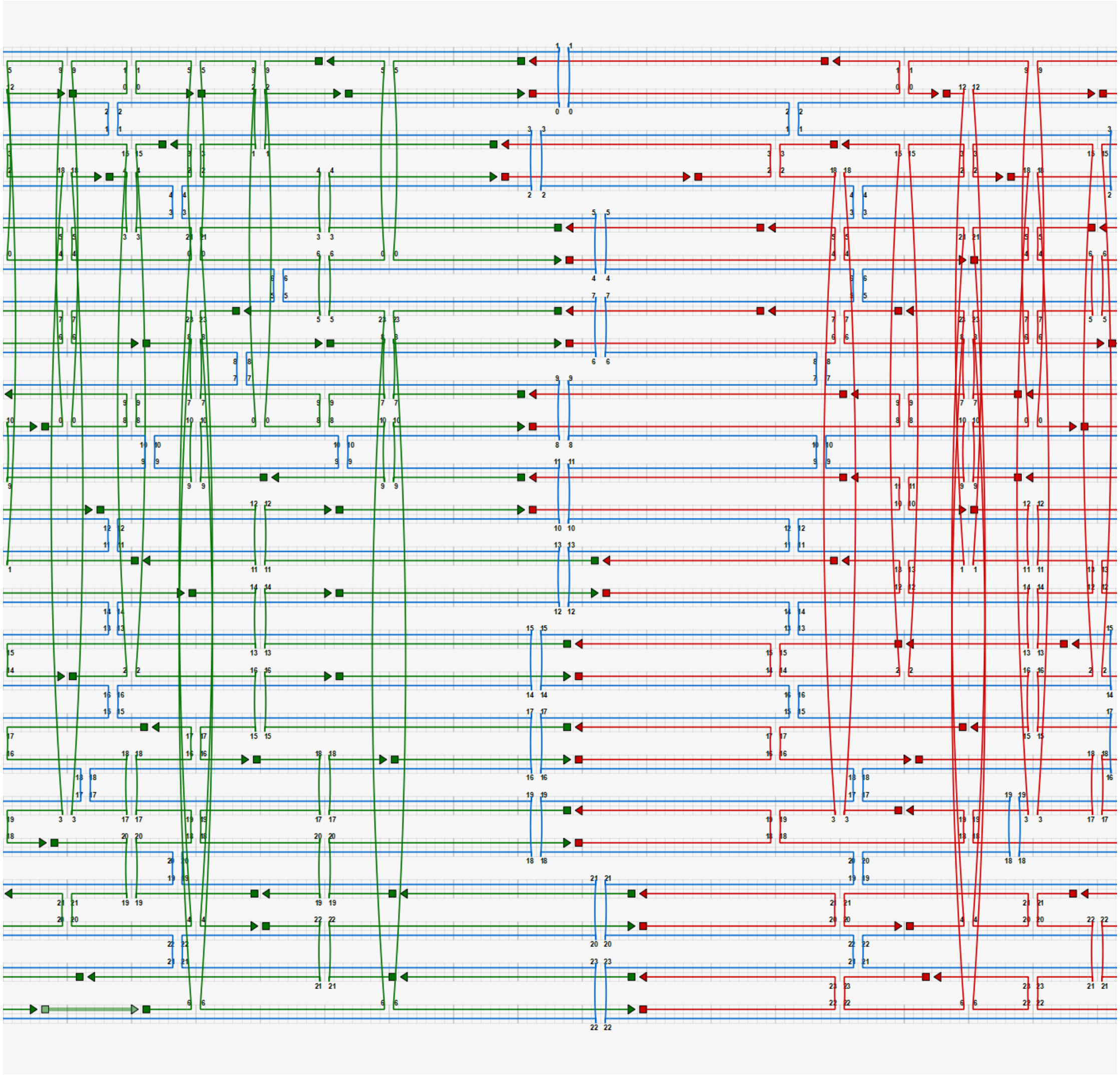

**Supplementary Figure 3: Cadnano design of 24 HB dimerization mechanism**

Dimerization technique depicted with the individual halves in green and red.

The Small Pitch Helix (SPH) was formed around a 48 helix bundle (48 HB) DNA origami trunk. This origami was synthesized with an 8064 nt scaffold strand and 205 staple strands. To prevent unintentional binding of NPs to the origamis ends, each side has 6 "crossbar" staples forming duplexes that connect helices in a horizontally fashion across the origami end. Gold nanoparticles can be attached to protruding handle strands depicted in red.

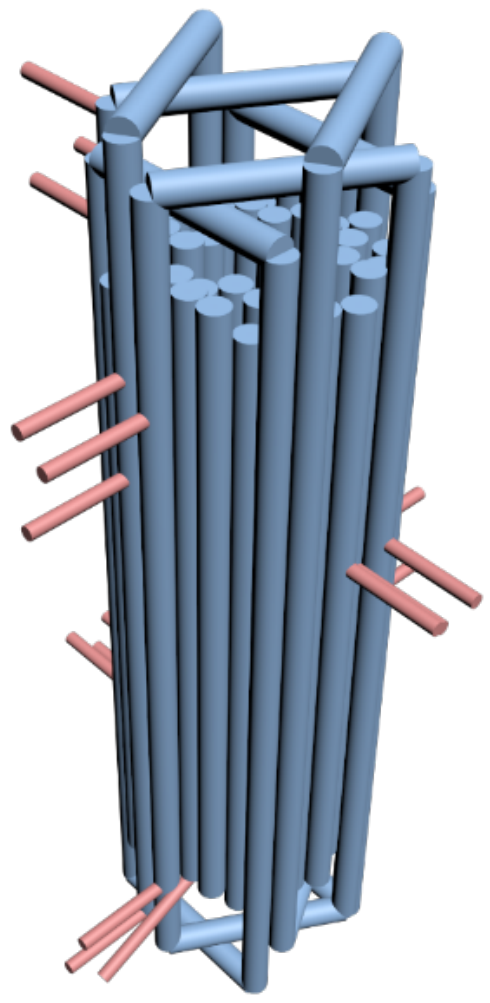

**Supplementary Figure 4: 3D Model of the DNA origami structure 48 HB**

48 HB with crossbar strands (cylinders represent DNA helices) and handles for nanoparticle attachment depicted in red.

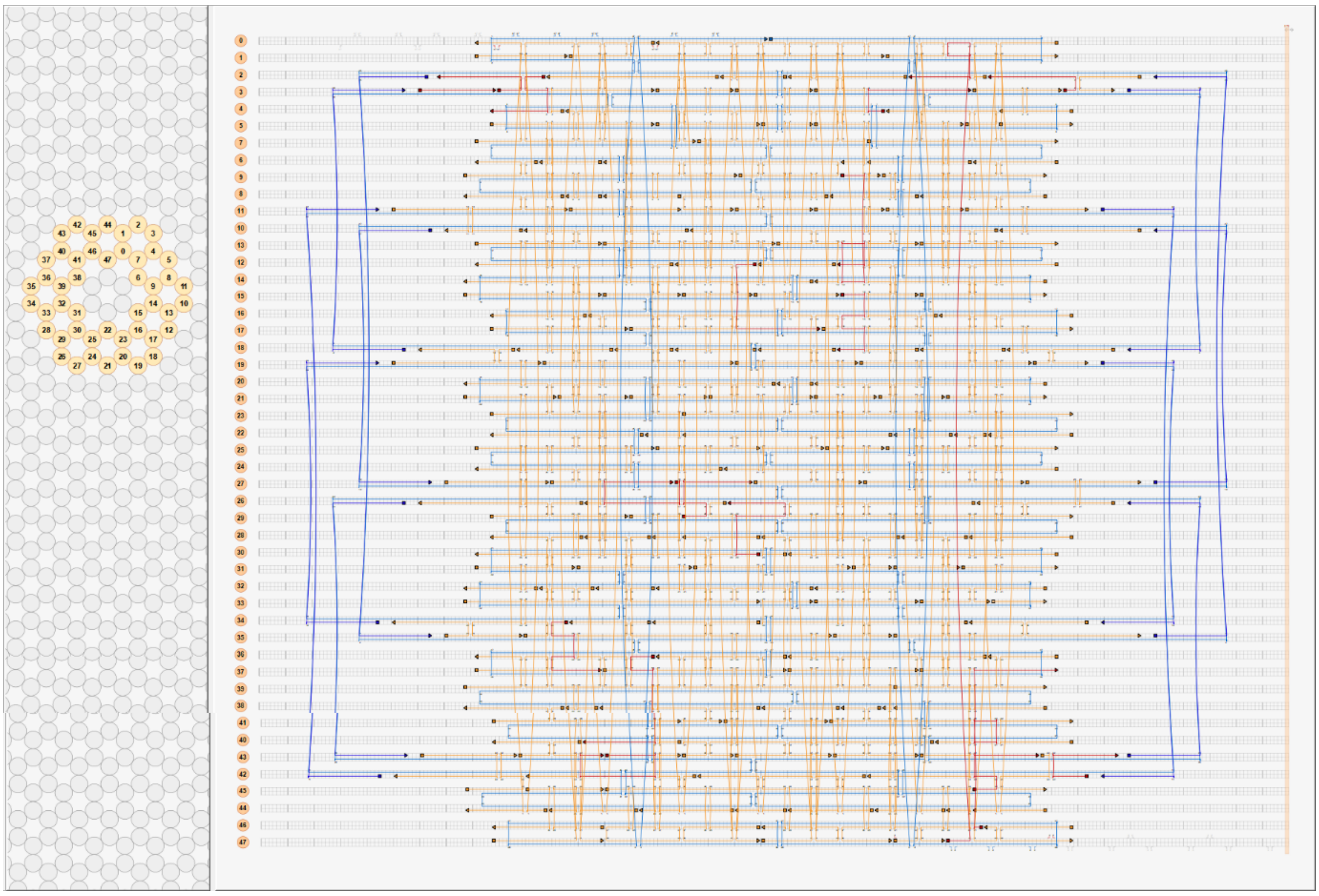

**Supplementary Figure 5: Cadnano design of the 48 HB monomer**

The scaffold strand is depicted in blue, crossbar staples strands are depicted in purple. Core staples are depicted in orange as well as the handle strands for the NPs in red.

## Gold particle helix parameters

After attachment of 40 nm spherical gold particles to the 24HB, the LPH forms a lefthanded helix with an offset of 29 nm and a rotation around the helical axis of 90° per particle. The radius of the helix amounts to 30 nm if measured from the center of particles to the center of the helix. The SPH also forms a lefthanded helix, however, with important geometric differences. The attached gold particles have a diameter of 30 nm, the offset per particle amounts to 11 nm and the radius, measured from the center of the particles to the center of the helix, is 30 nm.

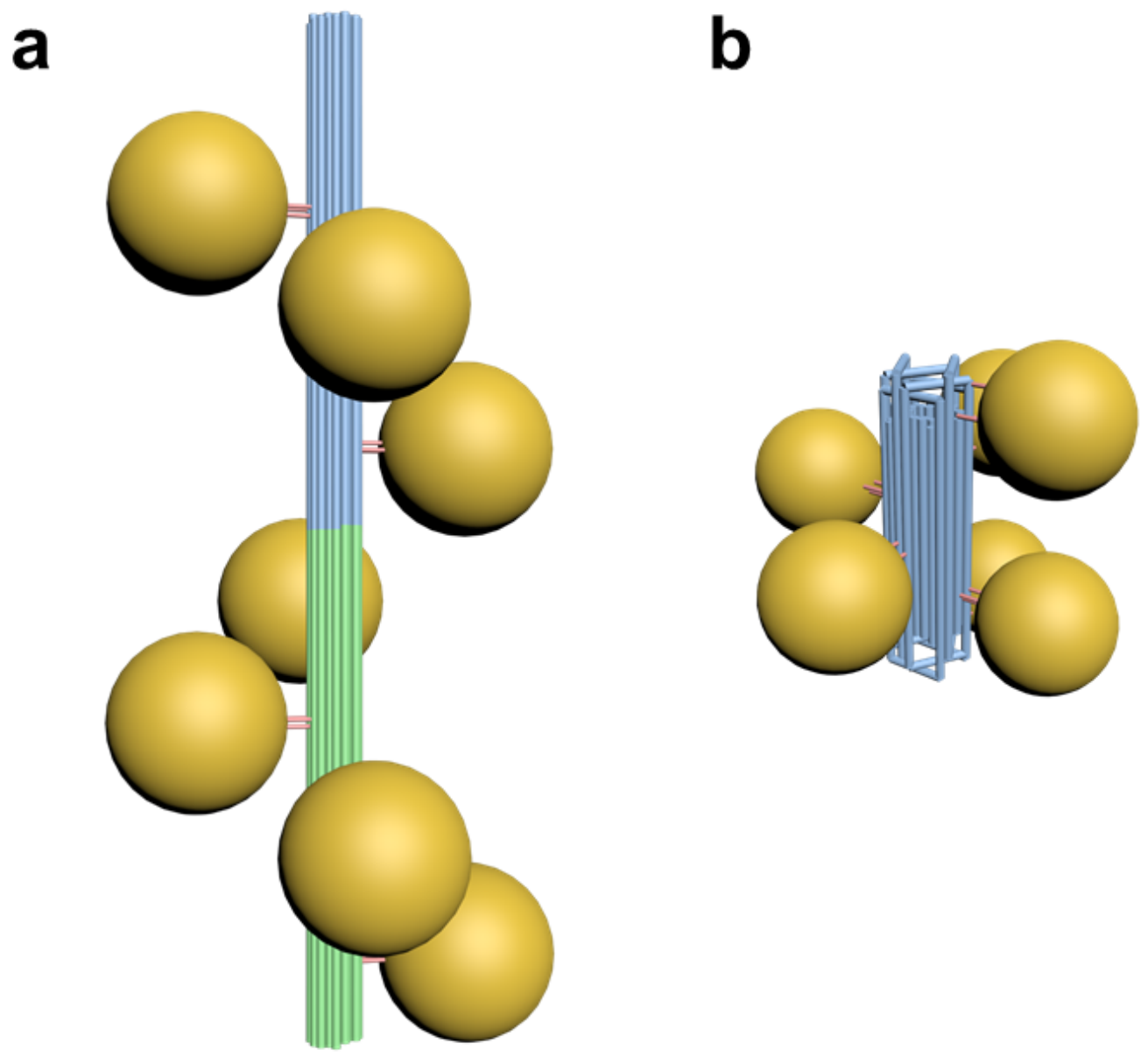

**Supplementary Figure 6: 3D Model of the complete nanohelix structures**

**a**, LPH dimer structure with individual halves depicted in blue and green and handles for nanoparticle attachment depicted in red. Gold spheres represent 40 nm spherical gold nanoparticles. **b**, SPH structure with 30 nm spherical gold nanoparticles.

## Supplementary Note 2: DNA Origami Synthesis

The 24 HB for the LPH was folded with a scaffold concentration of 16 nM and 160 nM core and endcap staples as well as 333 nM handle staples. For the dimer version 200 nM dimer staples were used. 1X TE and 24 mM $MgCl_2$ were added as buffer solution. The 48 HB for SPH was folded with a scaffold concentration of 12 nM and 120 nM core staples (including endcaps) as well as 600 nM handle staples. 1X TE and 26 mM $MgCl_2$ were added as buffer solution. Handles were added according to the desired number of gold nanoparticles to be attached in the next step.

The solutions containing the mixtures were heated up to 65°C and then cooled down to room temperature over the course of 24 h. Origamis were purified using gel electrophoresis with 0.7% agarose gel in a buffer of 1X TAE, 11 mM $MgCl_2$ and 0.05‰ Roti Stain as intercalating dye. The gel was run for 2.5 h at 70 V, before the origami dimer band was cut out under UV light and afterwards squeezed to redisperse the sample in buffer. For the dimer version, the left side of the origami and the right side were combined in equal amounts and left to dimerize over 24 h.

For TEM analysis, samples were incubated for 15 min on copper grids (Ted Pella Inc., Redding, USA) before being dabbed off with a filter paper and subsequently stained with 2% uranyl format in two steps. In the first step the uranyl format solution only quickly washes the grid, in the second step it is left to incubate for 15 s before being dabbed off. Images were taken with a JEOL JEM 1011 electron microscope at 80 kV.

**Supplementary Table 1: Monomer 24 HB DNA origami protocol**

| Component | Concentration | Amount | End Concentration |
|---|---|---|---|
| Scaffold 8064 | 100 nM | 16 µL | 16 nM |
| Core Staples | 320.5 nM | 49.9 µL | 160 nM |
| Endcap Staples L | 2083.3 nM | 9.6 µL | 160 nM |
| Endcap Staples R | 2380.9 nM | 8.4 µL | 160 nM |
| Handles | 16 666.7 nM | 2 µL each | 333 nM |
| TE | 20X | 5 µL | 1 X |
| MgCl2 | 1 M | 2.4 µL | 24 mM |
| H2O | - | 0.7-6.7 µL | - |
| **Total** | **100 nM** | **100 µL** | |

**Supplementary Table 2: Left side dimer 24 HB DNA origami protocol**

| Component | Concentration | Amount | End Concentration |
|---|---|---|---|
| Scaffold 8064 | 100 nM | 16 µL | 16 nM |
| Core Staples | 320.5 nM | 49.9 µL | 160 nM |
| Endcap Staples L | 2083.3 nM | 9.6 µL | 160 nM |
| Dimer Staples | 2631.6 nM | 7.6 µL | 200 nM |
| Handles | 16666.7 nM | 2 µL each | 333 nM |
| TE | 20X | 5 µL | 1 X |
| MgCl2 | 1 M | 2.4 µL | 24 mM |
| H2O | - | 1.5-7.5 µL | - |
| **Total** | **100 nM** | **100 µL** | |

**Supplementary Table 3: Right side dimere 24 HB DNA origami protocol**

| **Component** | **Concentration** | **Amount** | **End Concentration** |
|---|---|---|---|
| Scaffold 8064 | 100 nM | 16 µL | 16 nM |
| Core Staples | 320.5 nM | 49.9 µL | 160 nM |
| Endcap Staples R | 2380.9 nM | 8.4 µL | 160 nM |
| Dimer Staples | 2631.6 nM | 7.6 µL | 200 nM |
| Handles | 16666.7 nM | 2 µL each | 333 nM |
| TE | 20X | 5 µL | 1 X |
| MgCl2 | 1 M | 2.4 µL | 24 mM |
| H2O | - | 2.7-8.7 µL | - |
| **Total** | **100 nM** | **100 µL** | |

**Supplementary Table 4: Monomer 48 HB DNA origami protocol**

| **Component** | **Concentration** | **Amount** | **End Concentration** |
|---|---|---|---|
| Scaffold 8064 | 100 nM | 12 µL | 12 nM |
| Staples | 244 nM | 49.2 µL | 120 nM |
| Handles | 16666.7 nM | 3.6 µL each | 600 nM |
| TE | 20X | 5 µL | 1 X |
| MgCl2 | 1 M | 2.6 µL | 26 mM |
| H2O | - | 9.6-27.6 µL | - |
| **Total** | **100 nM** | **100 µL** | |

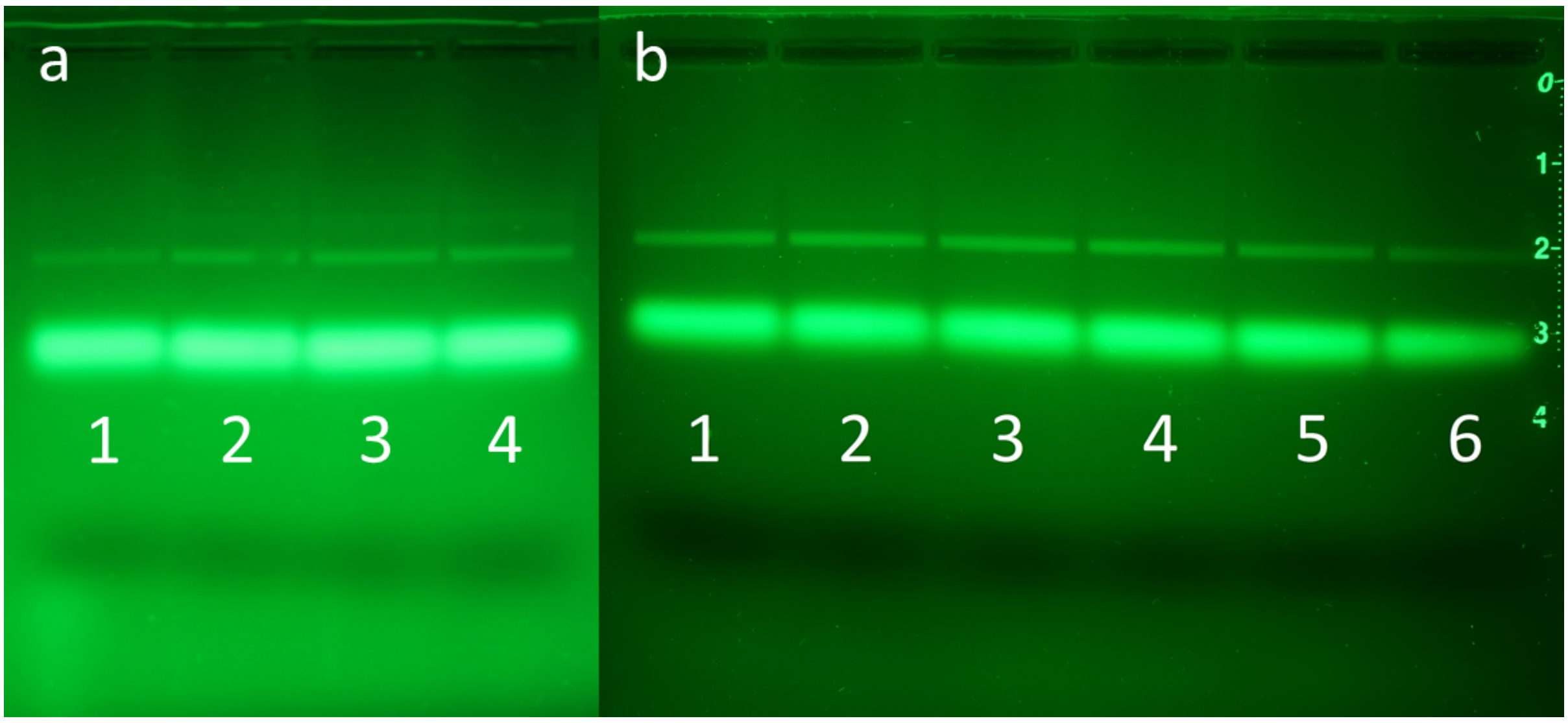

**Supplementary Figure 7: DNA origami gel electrophoresis bands**

Bands showing **a**, 24 HB monomers and **b**, 48 HB monomers with different numbers of handles for nanoparticles (1-4 for 24 HB and 1-6 for 48 HB). Gel electrophoresis was performed for 2.5 hours at 70 V, showing a band of DNA origami structure after folding (upper band) and a band of excess staples (lower band).

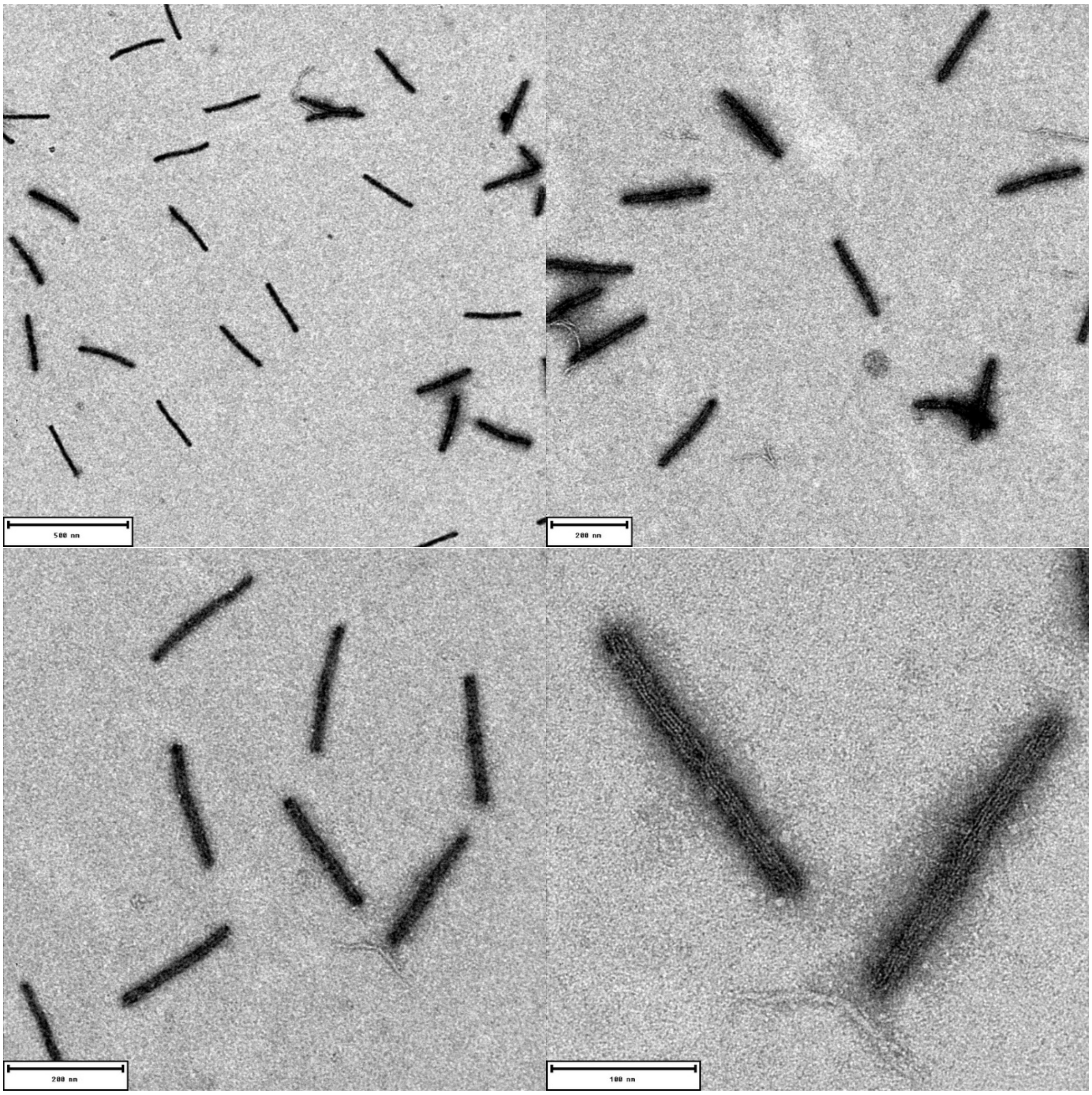

**Supplementary Figure 8: Electron micrographs of 24 HB dimers**

The DNA origami were purified using gel electrophoresis purification and imaged with transmission electron microscopy using uranyl format for staining. Scale bars: 500 nm, 200 nm, 200 nm and 100 nm

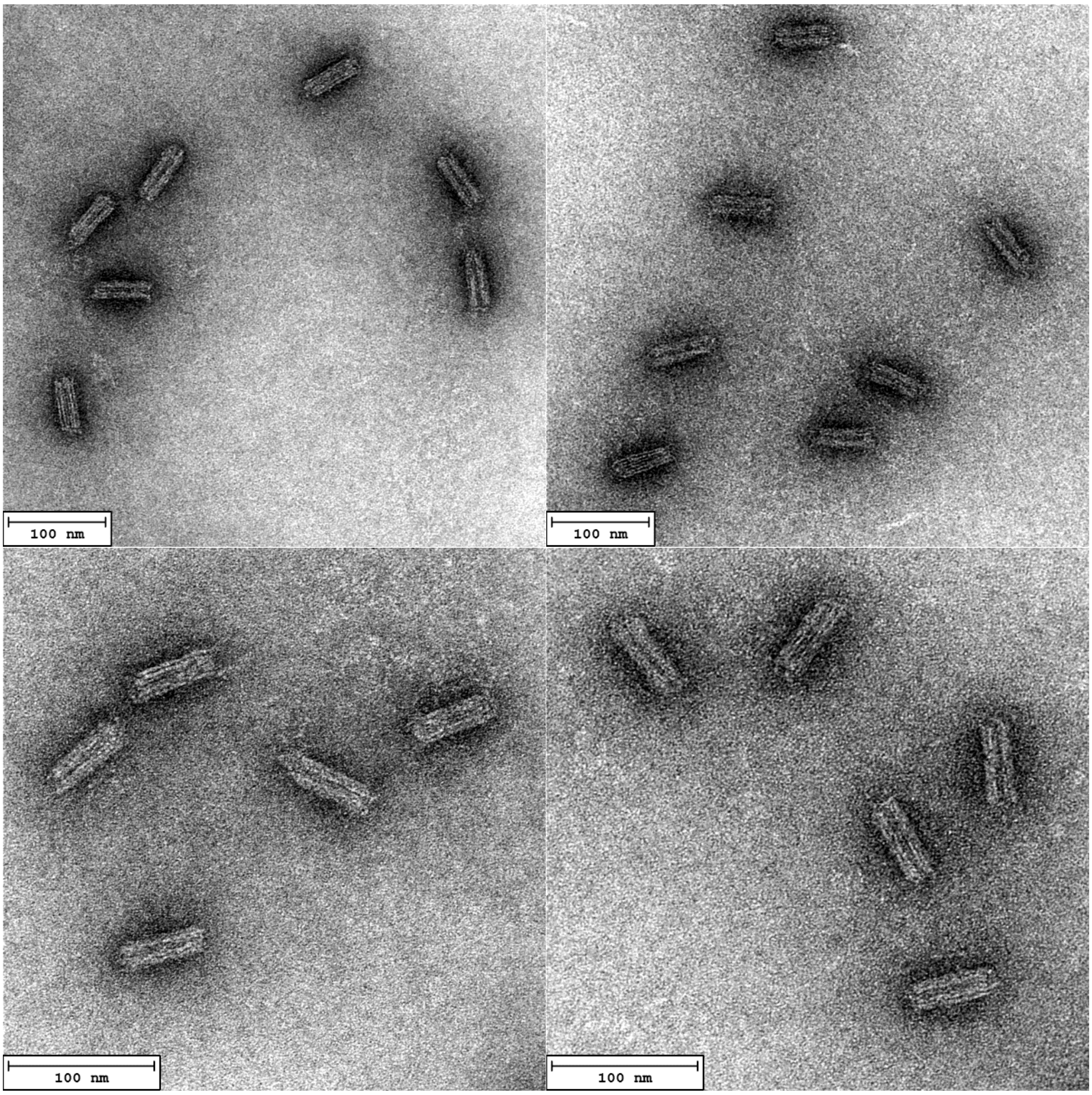

**Supplementary Figure 9: Electron micrographs of 48 HB monomers**

The DNA origami were purified using gel electrophoresis purification and imaged with transmission electron microscopy using uranyl format for staining. Scale bar: 100 nm.

# Supplementary Note 3: DNA Origami-Nanoparticles Assembly Synthesis

30 nm (SPH) and 40 nm (LPH) gold nanoparticles were incubated at an optical density (OD) of 4 with 10 mM thiol-modified DNA oligonucleotides, previously activated with TCEP, and 0.02% SDS. Samples were frozen, thawed and purified using gel electrophoresis with a 0.7% agarose gel in a buffer of 1X TAE, 11 mM $MgCl_2$, run for 1.5 h at 120 V. Subsequently the correct monomer bands were cut and squeezed to redisperse in buffer.

For the synthesis of the LPH and SPH samples, gold nanoparticles were added to the origami structures in a ratio of 5:1 for each particle in a buffer of 1X TAE, 11 mM MgCl2 plus 500 mM NaCl and incubated for between 1 and 24 h. The samples were purified using gel electrophoresis with a 1.5% agarose gel in a buffer of 1X TAE, 11 mM $MgCl_2$, run for 4-6 h at 70 V. The different bands of fully formed structures were cut out and squeezed and analyzed by TEM.

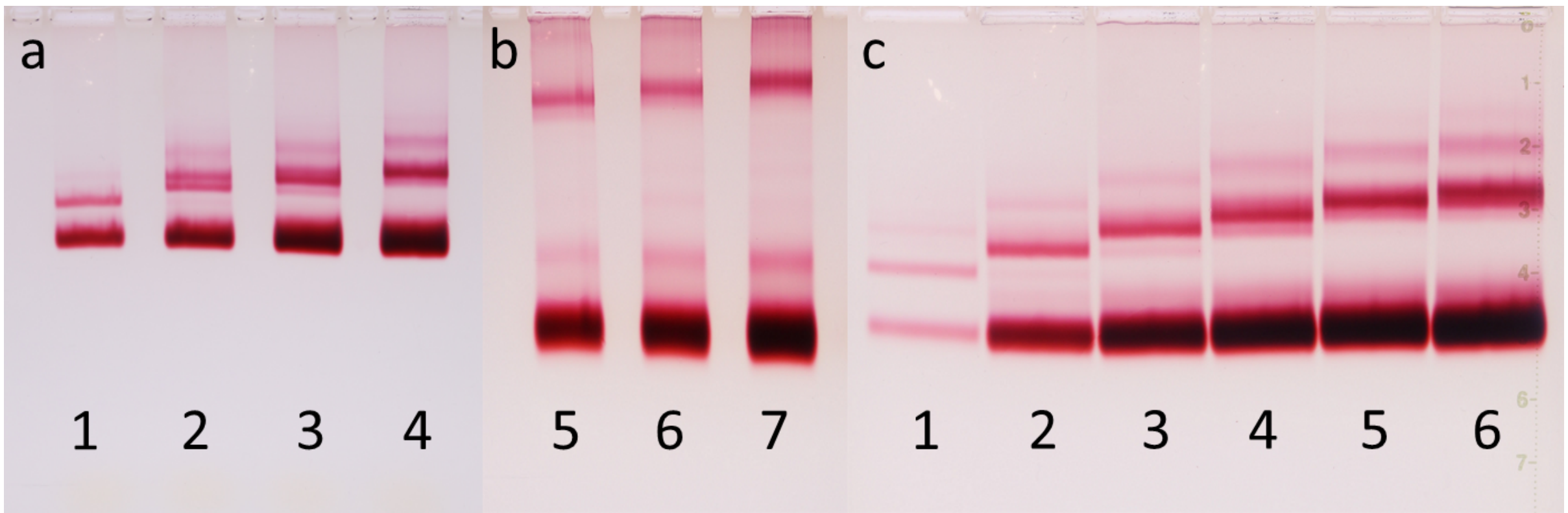

**Supplementary Figure 10: Nanostructure gel electrophoresis bands**

Bands showing **a**, LPH-i to LPH-iv **b**, LPH-v to LPH-vii and **c**, SPH-i to SPH-vi. Gel electrophoresis was performed for 3 hours (a), 6 hours (b) and 3 hours (c) at 70 V, showing bands of origami structures with different numbers of gold nanoparticles attached (upper bands) and a thicker band of excessive gold nanoparticles (lower band).

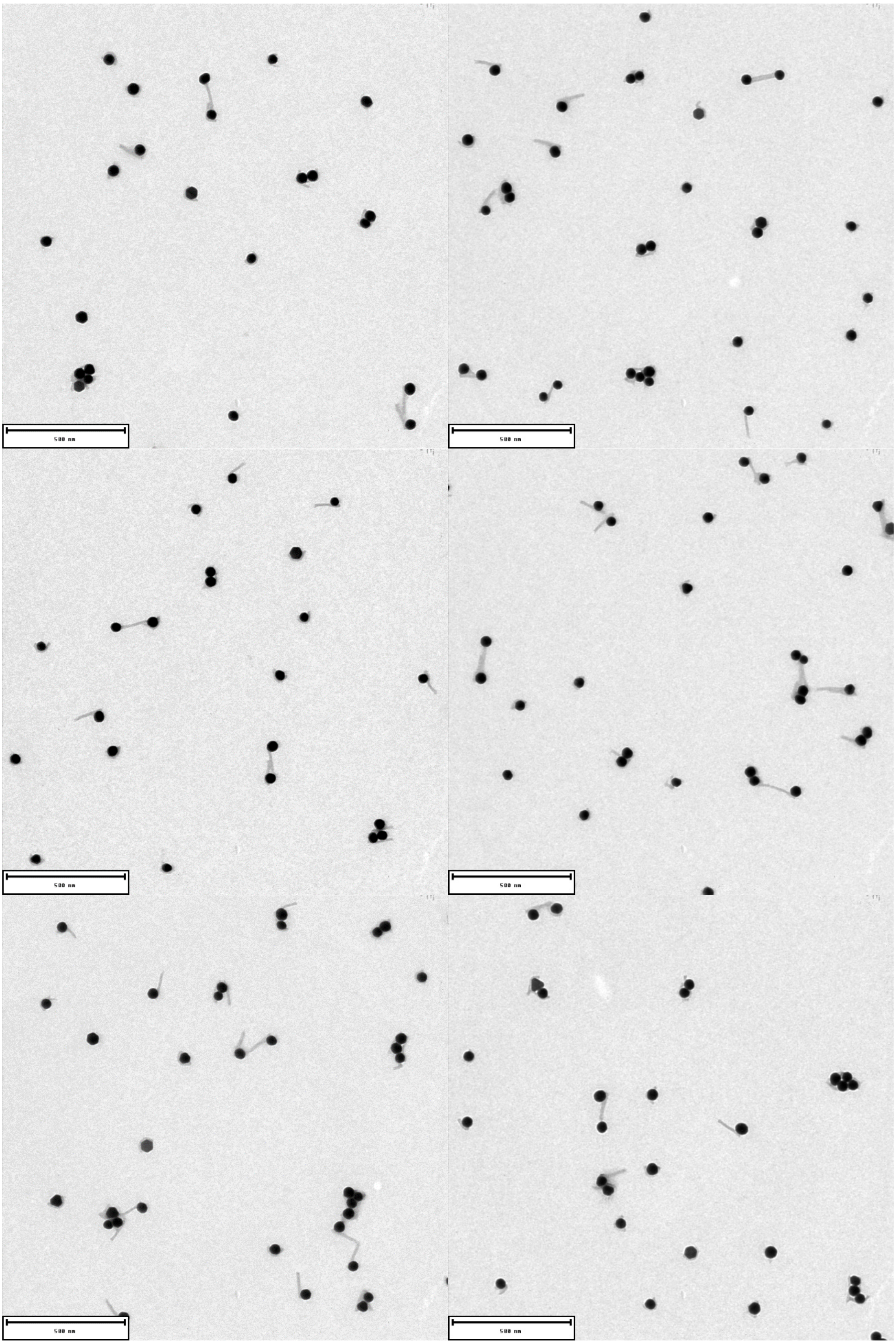

**Supplementary Figure 11: Electron micrographs of LPH-i**

The structures were purified using gel electrophoresis purification and imaged with transmission electron microscopy using uranyl format for staining. Scale bar: 500nm.

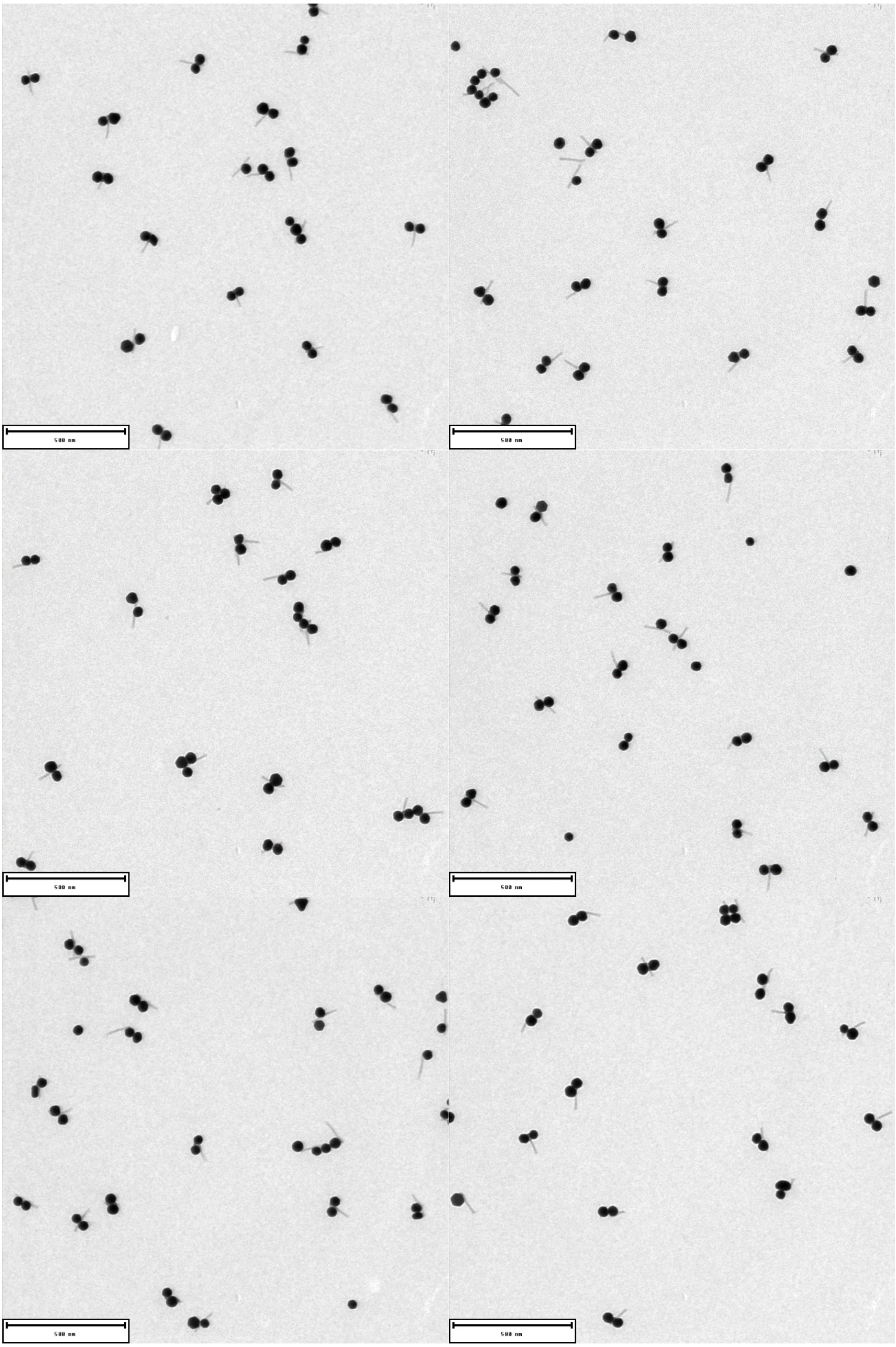

**Supplementary Figure 12: Electron micrographs of LPH-ii**

The structures were purified using gel electrophoresis purification and imaged with transmission electron microscopy using uranyl format for staining. Scale bar: 500nm.

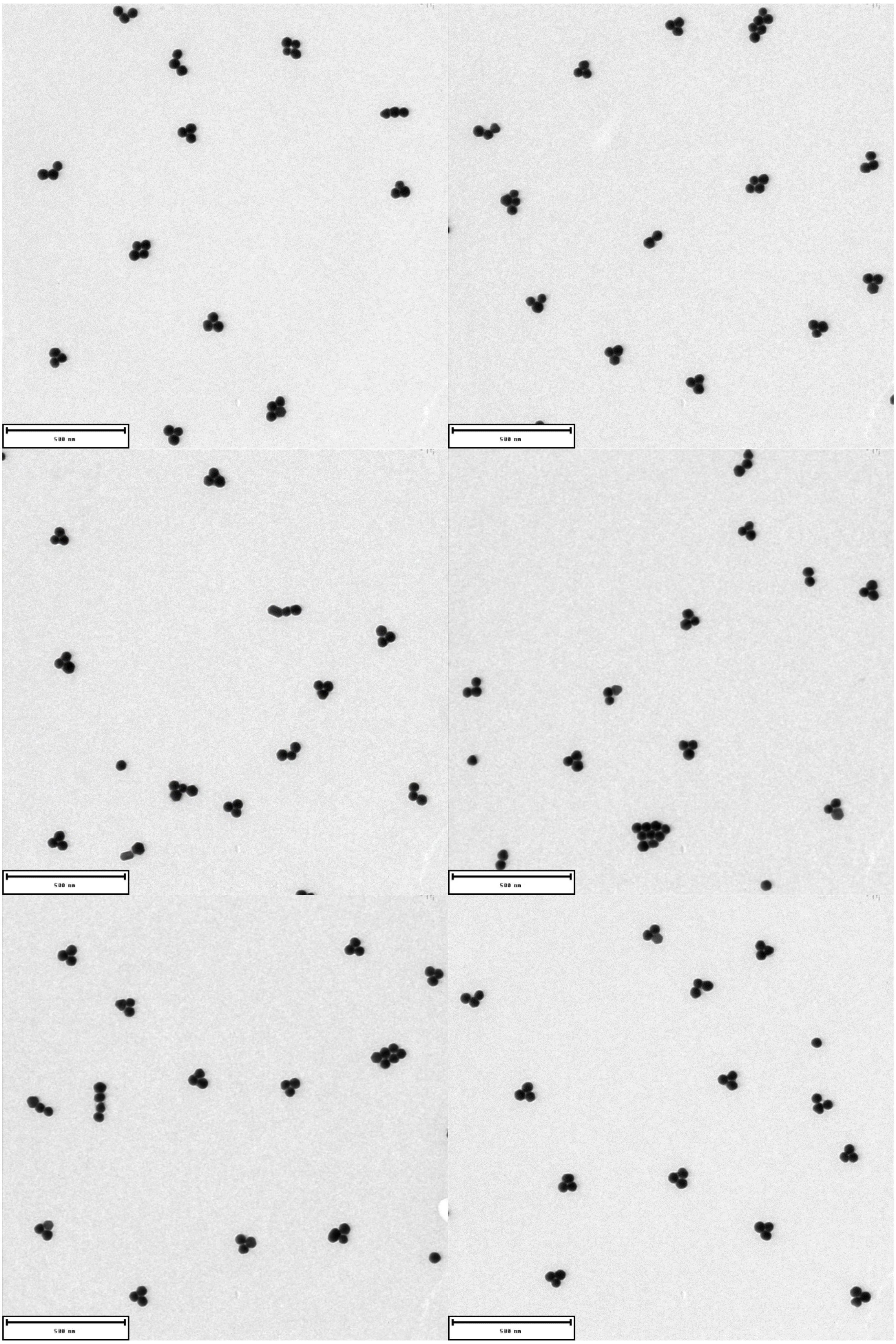

**Supplementary Figure 13: Electron micrographs of LPH-iii**

The structures were purified using gel electrophoresis purification and imaged with transmission electron microscopy using uranyl format for staining. Scale bar: 500nm.

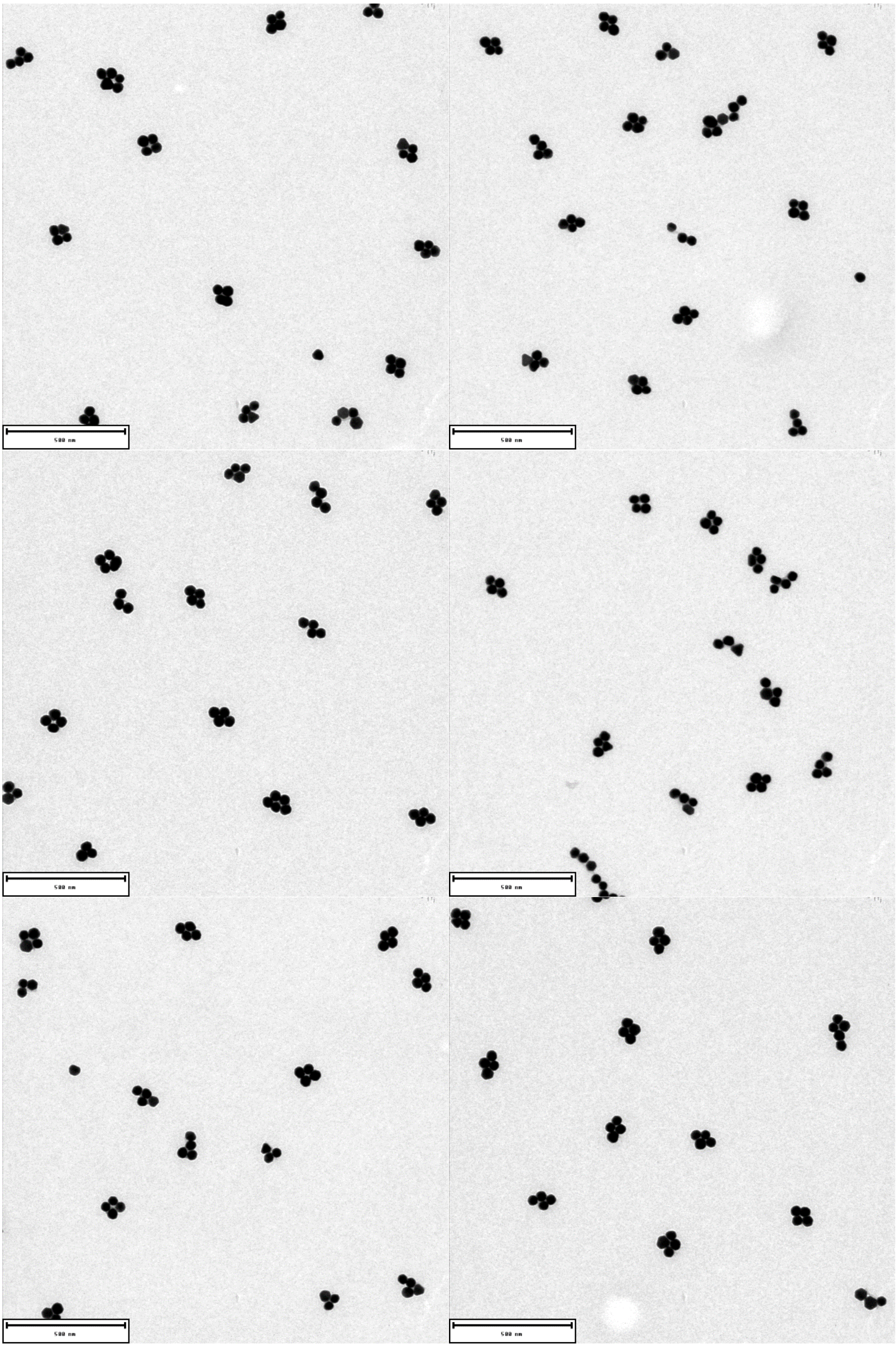

**Supplementary Figure 14: Electron micrographs of LPH-iv**

The structures were purified using gel electrophoresis purification and imaged with transmission electron microscopy using uranyl format for staining. Scale bar: 500nm.

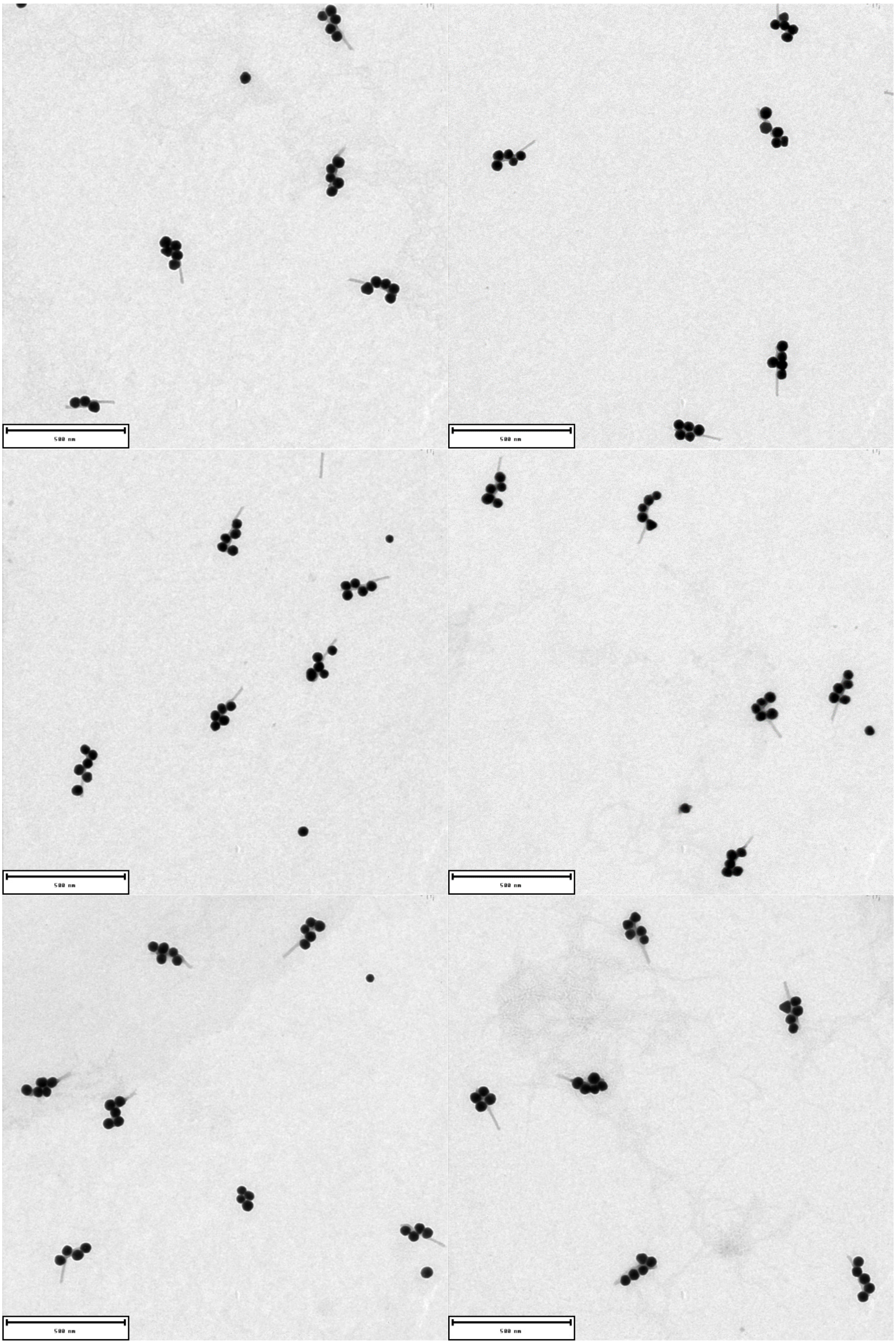

**Supplementary Figure 15: Electron micrographs of LPH-v**

The structures were purified using gel electrophoresis purification and imaged with transmission electron microscopy using uranyl format for staining. Scale bar: 500nm.

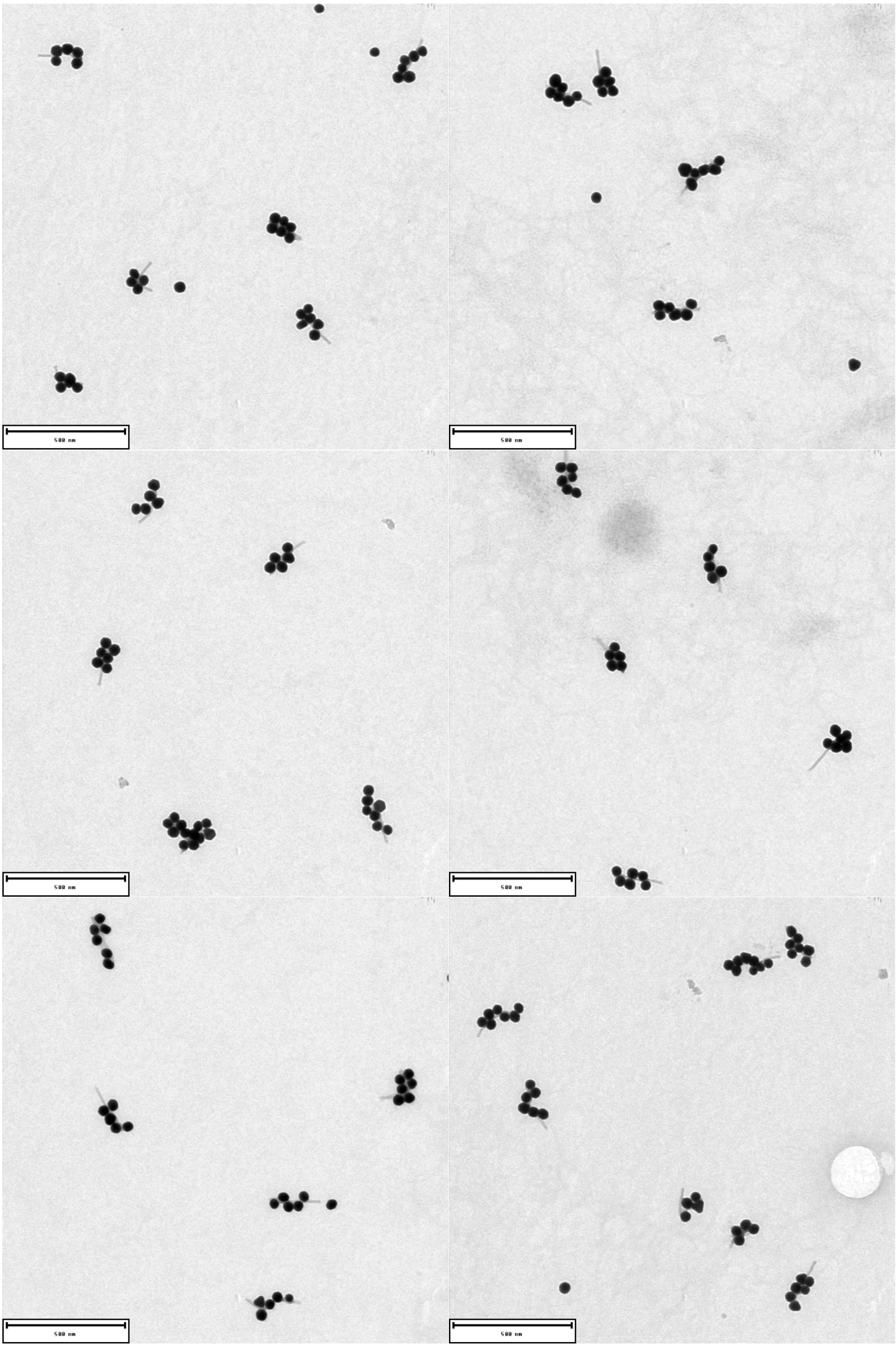

**Supplementary Figure 16: Electron micrographs of LPH-vi**

The structures were purified using gel electrophoresis purification and imaged with transmission electron microscopy using uranyl format for staining. Scale bar: 500nm.

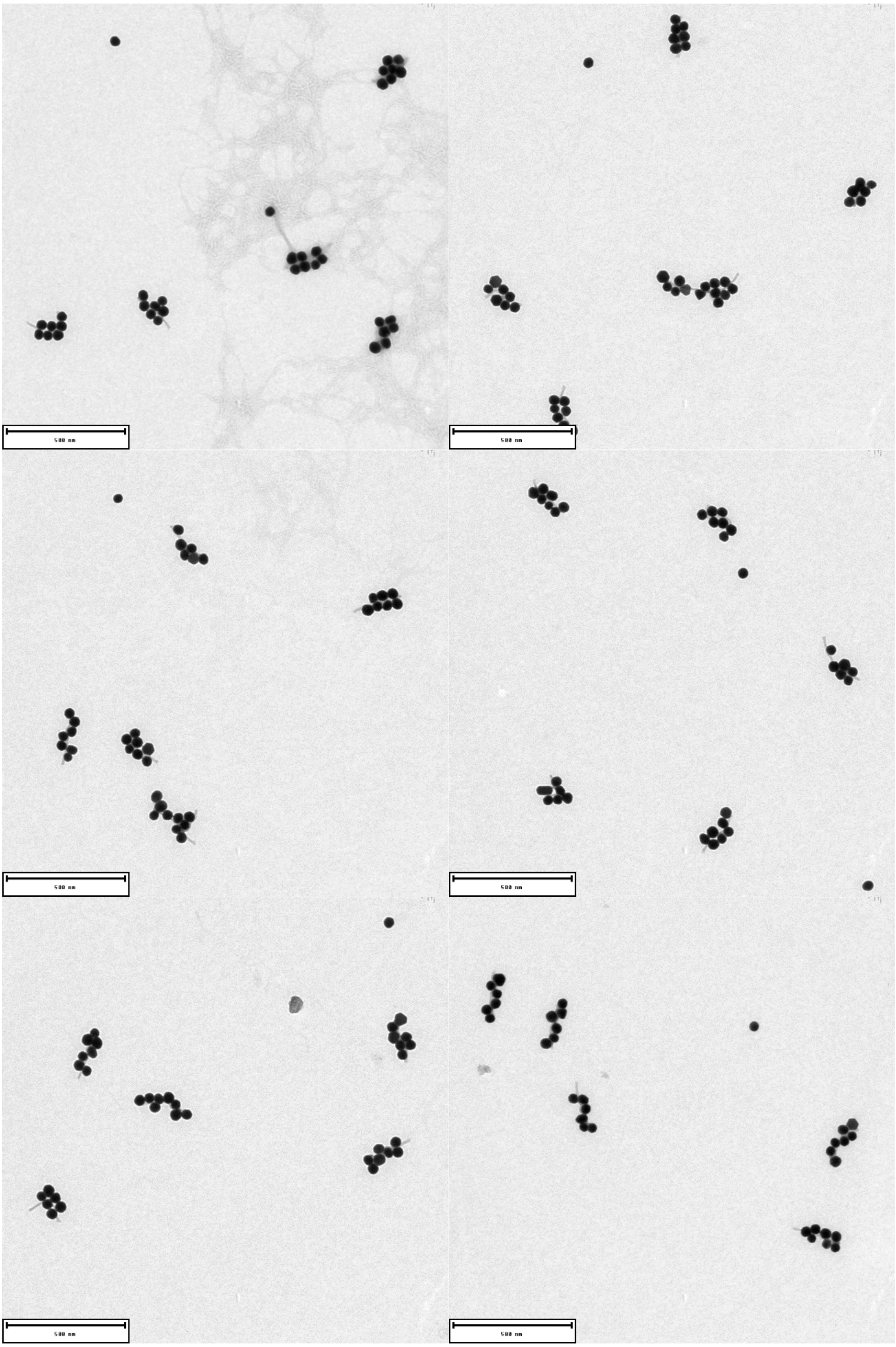

**Supplementary Figure 17: Electron micrographs of LPH-vii**

The structures were purified using gel electrophoresis purification and imaged with transmission electron microscopy using uranyl format for staining. Scale bar: 500nm.

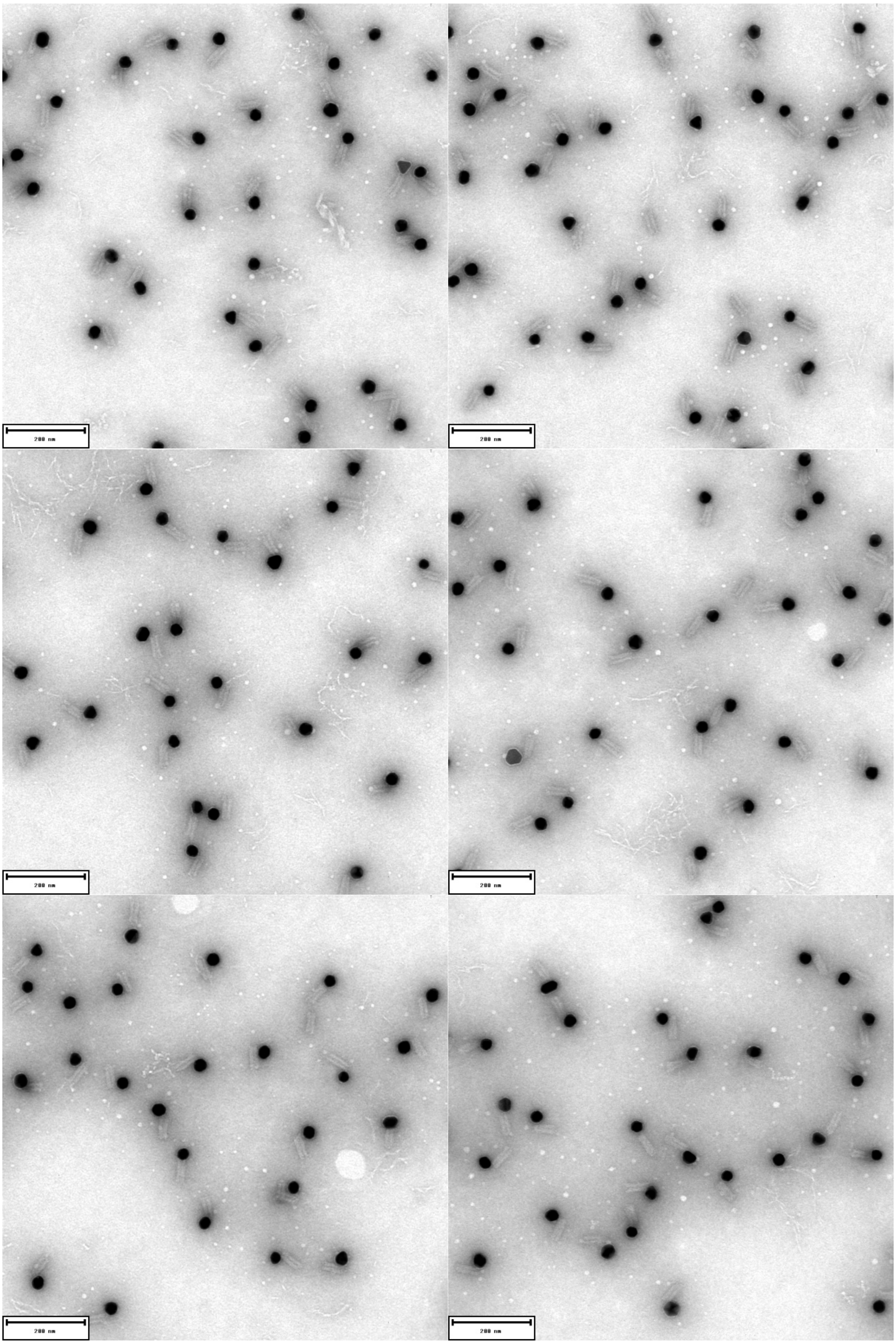

**Supplementary Figure 18: Electron micrographs of SPH-i**

The structures were purified using gel electrophoresis purification and imaged with transmission electron microscopy using uranyl format for staining. Scale bar: 200 nm.

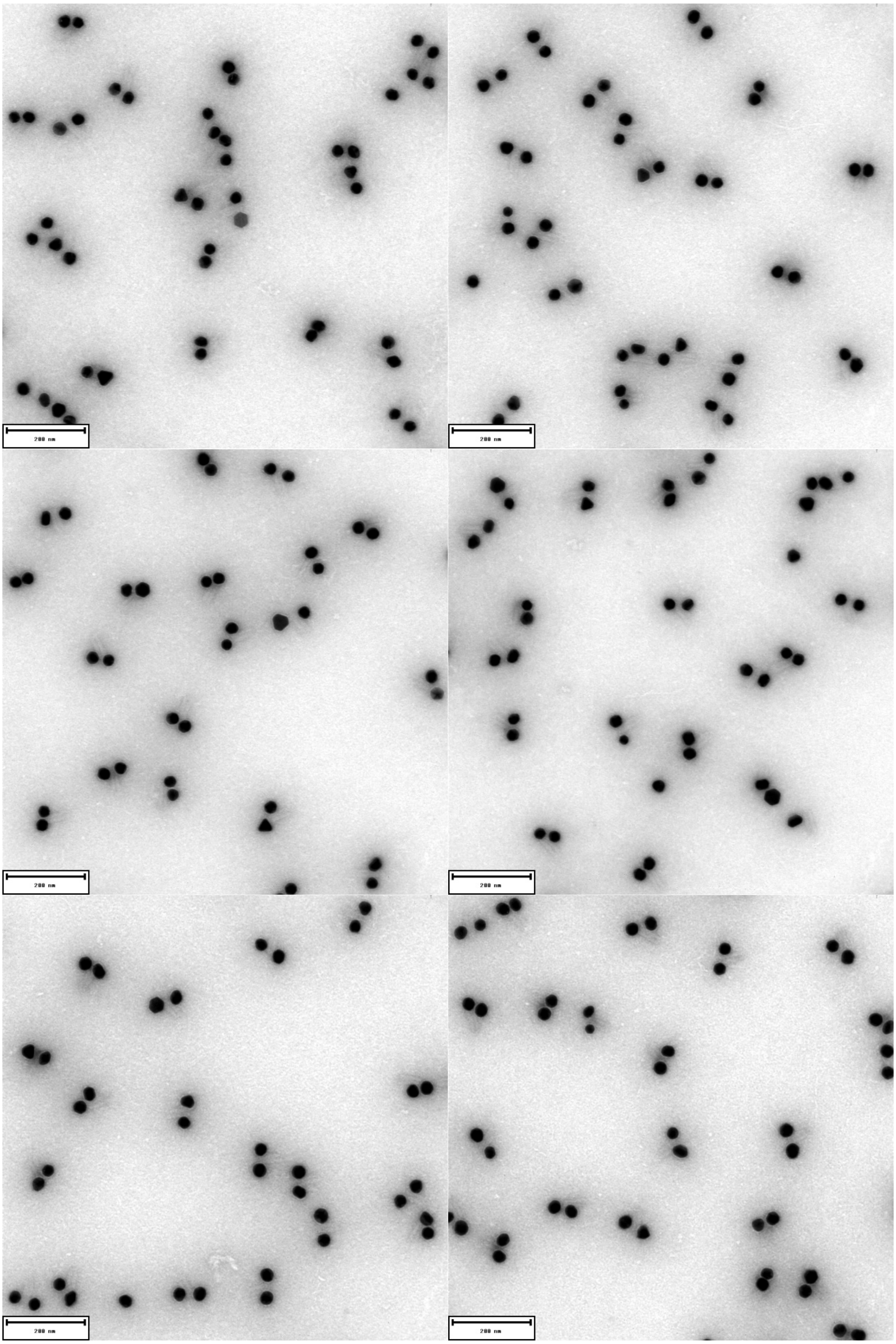

**Supplementary Figure 19: Electron micrographs of SPH-ii**

The structures were purified using gel electrophoresis purification and imaged with transmission electron microscopy using uranyl format for staining. Scale bar: 200 nm.

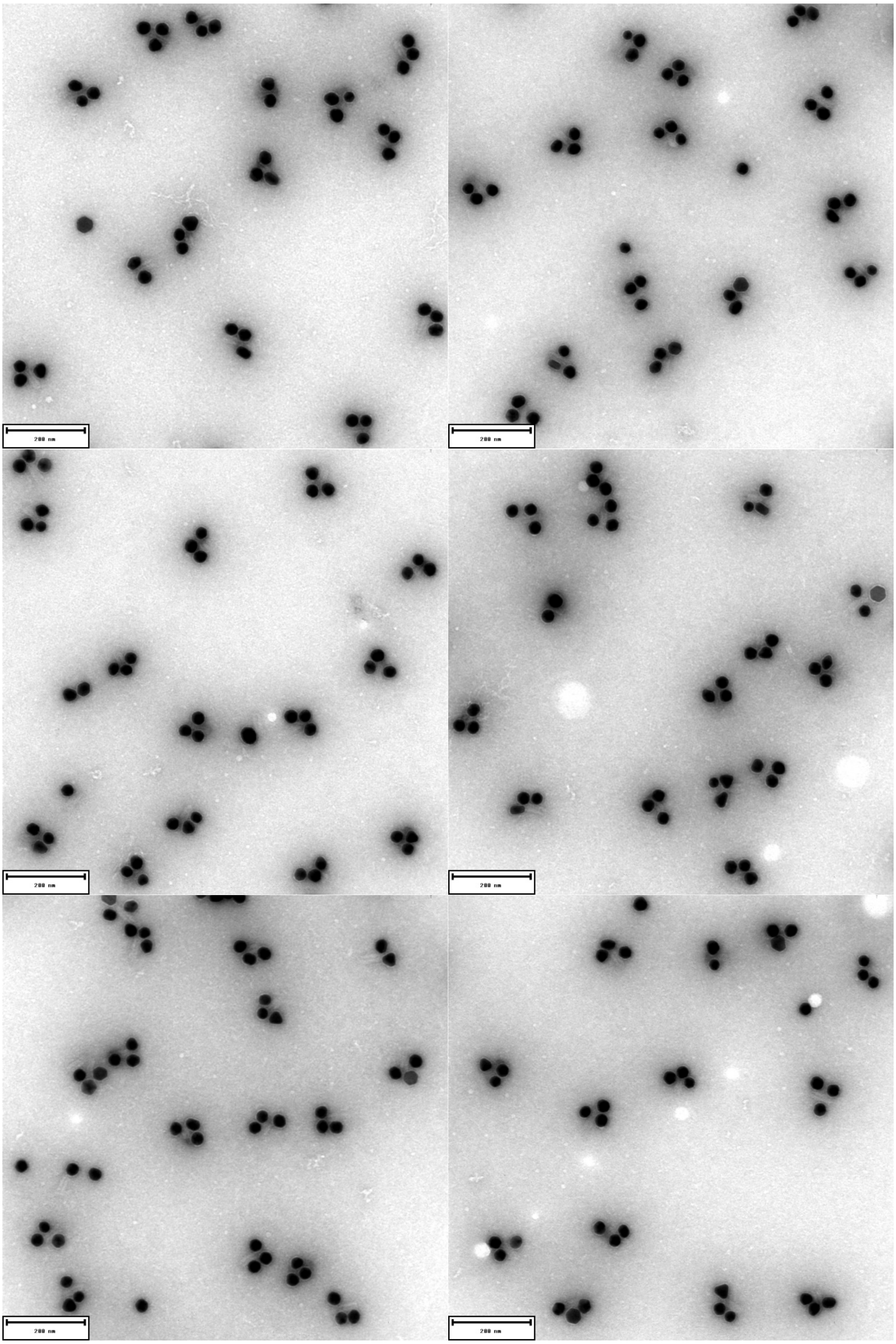

**Supplementary Figure 20: Electron micrographs of SPH-iii**

The structures were purified using gel electrophoresis purification and imaged with transmission electron microscopy using uranyl format for staining. Scale bar: 200 nm.

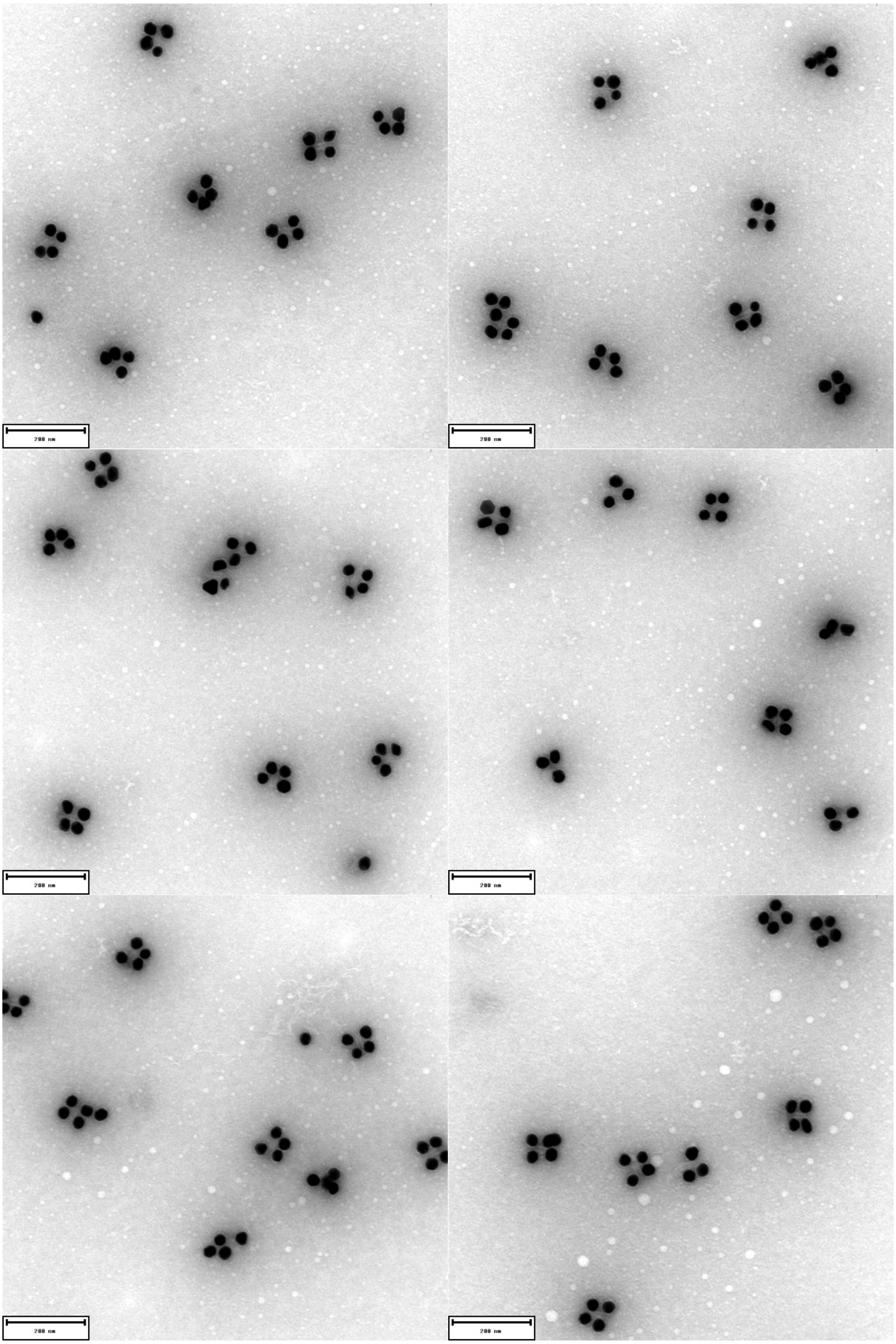

**Supplementary Figure 21: Electron micrographs of SPH-iv**

The structures were purified using gel electrophoresis purification and imaged with transmission electron microscopy using uranyl format for staining. Scale bar: 200 nm.

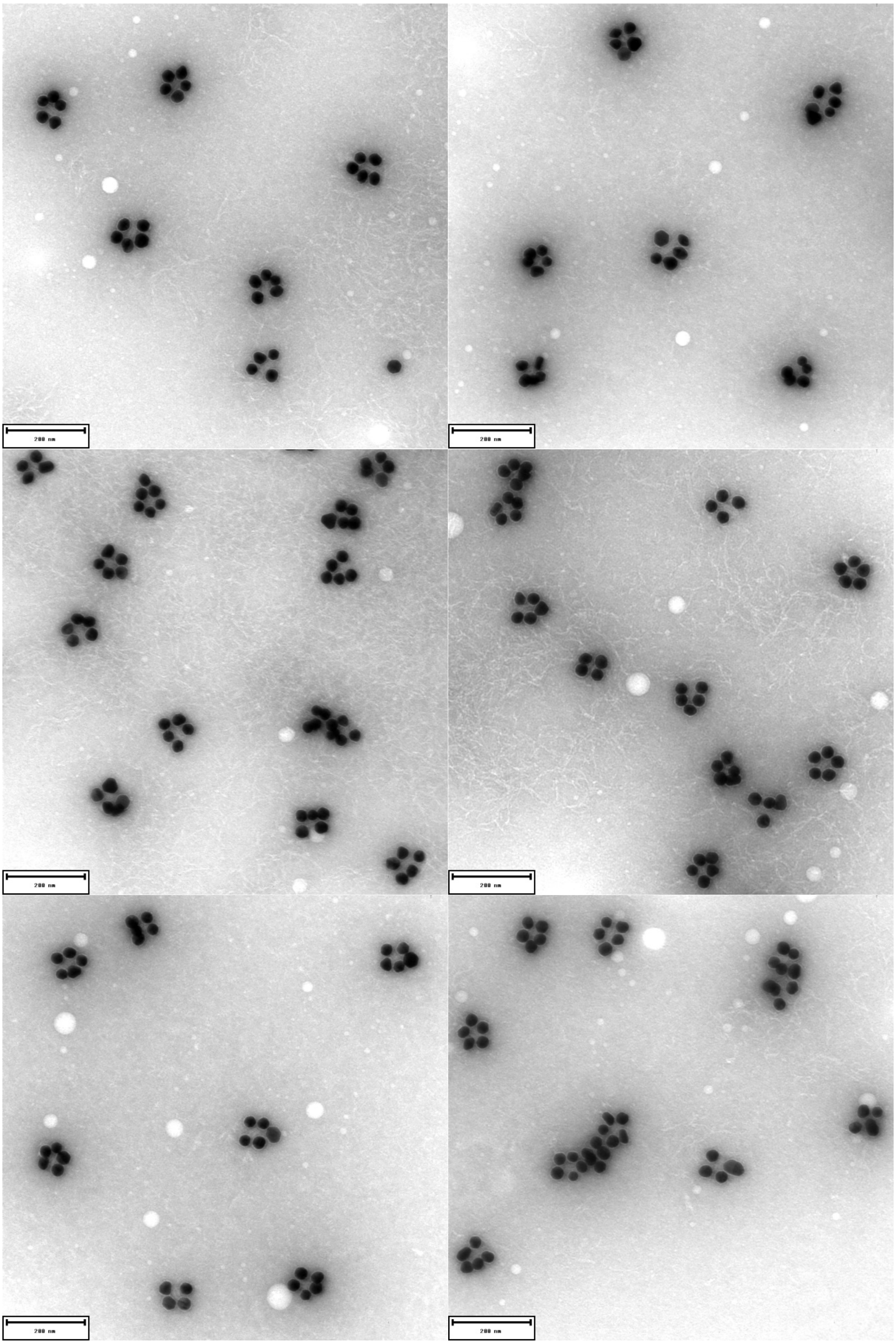

**Supplementary Figure 22: Electron micrographs of SPH-v**

The structures were purified using gel electrophoresis purification and imaged with transmission electron microscopy using uranyl format for staining. Scale bar: 200 nm.

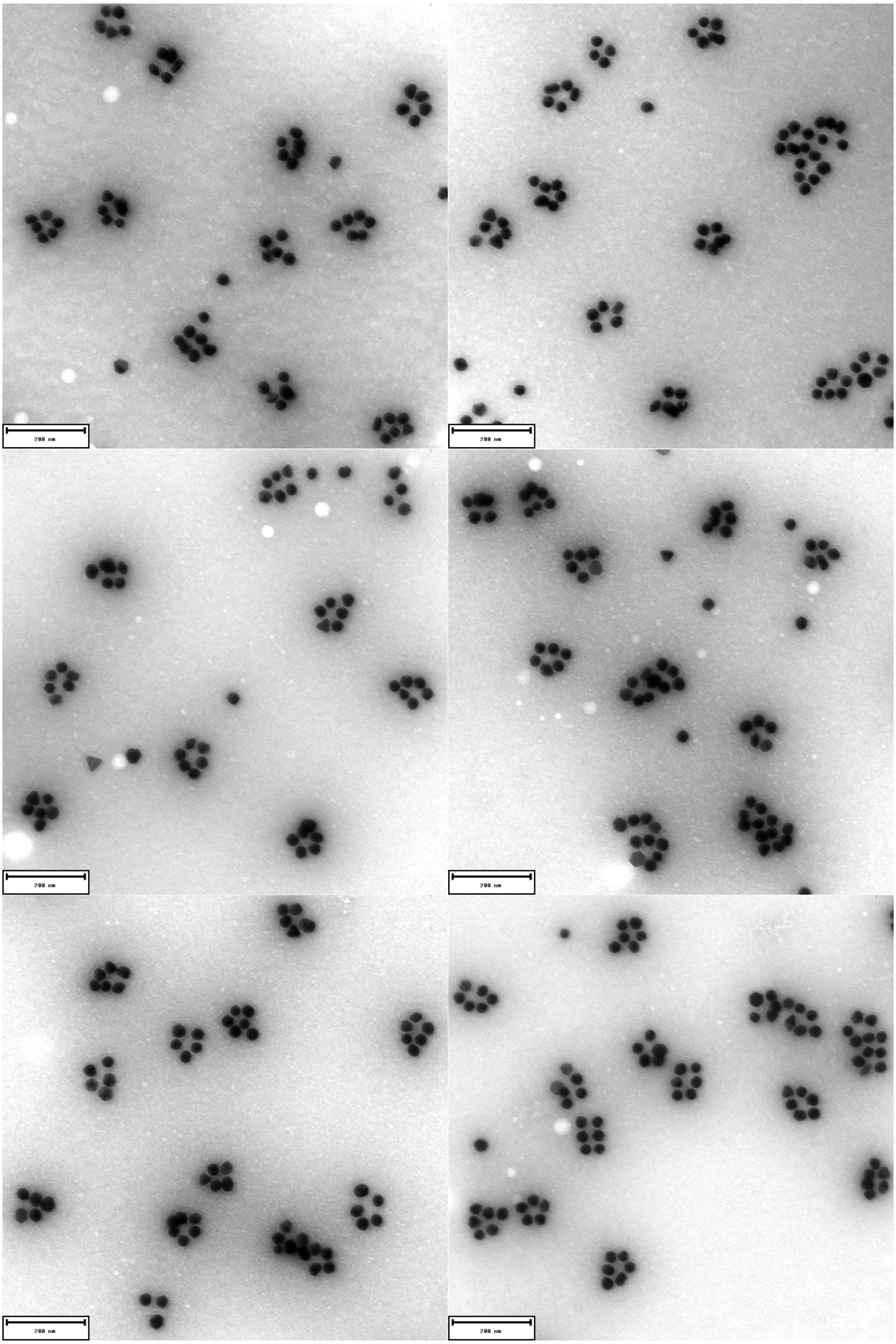

**Supplementary Figure 23: Electron micrographs of SPH-vi**

The structures were purified using gel electrophoresis purification and imaged with transmission electron microscopy using uranyl format for staining. Scale bar: 200 nm.

**Supplementary Table 5: LPH Synthesis Assembly Statistics**

| Number of particles | LPH-i | LPH-ii | LPH-iii | LPH-iv | LPH-v | LPH-vi | LPH-vii |
|---|---|---|---|---|---|---|---|
| **1** | 240 | 85 | 20 | 24 | 34 | 63 | 36 |
| **2** | 56 | 218 | 17 | - | 1 | 7 | 1 |
| **3** | 11 | 19 | 161 | 44 | 3 | 2 | 1 |
| **4** | 6 | 9 | 16 | 194 | 23 | 12 | 3 |
| **5** | 1 | 1 | 1 | 21 | 153 | 34 | 11 |
| **6** | - | 1 | 16 | 3 | 17 | 125 | 67 |
| **7** | - | 1 | - | 6 | - | 40 | 131 |
| **8** | - | - | 3 | 9 | - | 1 | 17 |
| **9** | - | - | 2 | - | 1 | 1 | 1 |
| **Total** | 314 | 334 | 236 | 277 | 196 | 222 | 232 |

**Supplementary Table 6: SPH Synthesis Assembly Statistics**

| Number of particles | SPH-i | SPH-ii | SPH-iii | SPH-iv | SPH-v | SPH-vi |
|---|---|---|---|---|---|---|
| **1** | 251 | 8 | 12 | 4 | 6 | 3 |
| **2** | - | 300 | 25 | 1 | 1 | 2 |
| **3** | - | 3 | 278 | 48 | 1 | 4 |
| **4** | - | 2 | 4 | 179 | 55 | 14 |
| **5** | - | 1 | 1 | 6 | 164 | 99 |
| **6** | - | - | - | 2 | 6 | 143 |
| **7** | - | - | - | 2 | - | 5 |
| **Total** | 251 | 314 | 320 | 242 | 233 | 270 |

## Supplementary Note 4: CD and Extinction Measurements

Samples were measured with a Chirascan circular dichroism spectrometer (Applied Photophysics, Surrey, UK) in cuvettes with 3 mm pathlengths. Spectra were collected in 1 nm steps with 1 s for each step. 2 measurements of two independent sets of samples were made and averaged for the LPH samples, 3 measurements of three independent sets of samples were made and averaged for the SPH samples.

## Supplementary Note 5: Numerical Simulations

In order to compute absorption and circular dichroism (CD) spectra for a given nanoassembly sample, we numerically solve the linear Maxwell's equations in the frequency domain, i.e., the steady-state field oscillating at a fixed optical frequency is computed. A geometry model is defined from the experimentally given parameters of a single nanoparticle helix arrangement (number and diameter of Au particles, their placement in space relatively to the DNA origami structure). The DNA origami is modelled as a dielectric cylinder of length 232nm (69nm), diameter 17nm (18nm), top and bottom edge rounding of 2nm for the LPH (SPH) assembly. For the SPH helix the cylinder is hollow with an inner diameter of 6nm. The model assumes wavelength dependent [Johnson&Christy] optical material parameters (refractive index) for gold, a constant refractive index of 1.58 for the dielectric cylinder modelling the DNA origami, and a constant refractive index of 1.33 for the background solution. In order to model the random orientation of the nanoassemblies in the background solution, the single assembly in our model is illuminated using left- and right-hand circularly polarized plane waves from 60 directions, and the results are obtained by averaging over the independent solutions for the various illuminations.

For computing the scattered light field ("near field") we use an adaptive finite-element method, implemented in the solver JCMsuite, version V4.2. The geometry model is discretized using tetrahedral mesh elements. Typical mesh element side lengths range between 2.5nm and 20nm. The electric field solution is discretized using second-order edge-elements. For checking the sufficient accuracy of the achieved results, reference solutions with third-order edge-elements have been computed for the same space discretizations. Transparent boundary conditions are realized by using perfectly matched layers.

The absorbed electromagnetic field energy for each source is calculated through volume integration of the electromagnetic field energy density in the absorbing Au nanoparticles. Correspondingly, the scattered electromagnetic field energy is calculated through surface integration of the electromagnetic energy flux density over the boundary of the computational domain. Absorption and CD signals are obtained from these quantities. For computing the spectral responses of the various nanoparticle assemblies, the physical quantities in the input for the simulation software are parameterized, and the FEM computations are distributed automatically to various threads on a workstation, for parallel execution.

For obtaining field patterns for visualization purposes (Fig 4), for each source direction and polarization, the electric field energy density is computed in a post-process. The densities are summed up to give the total energy density field in a further post-process. This total field is then exported on the geometry patches of the material interfaces within the computational domain, and displayed using a field viewer.

**Supplementary References**